\documentclass[letterpaper,twocolumn,english,showpacs,preprintnumbers,amsmath,amssymb,superscriptaddress,nofootinbib,prl]{revtex4}
\usepackage[T1]{fontenc}
\usepackage[latin9]{inputenc}
\usepackage{verbatim}
\usepackage{float}
\usepackage{amsmath}
\usepackage{graphicx}
\usepackage{setspace}

\makeatletter

\pdfpageheight\paperheight
\pdfpagewidth\paperwidth

\@ifundefined{textcolor}{}
{%
 \definecolor{BLACK}{gray}{0}
 \definecolor{WHITE}{gray}{1}
 \definecolor{RED}{rgb}{1,0,0}
 \definecolor{GREEN}{rgb}{0,1,0}
 \definecolor{BLUE}{rgb}{0,0,1}
 \definecolor{CYAN}{cmyk}{1,0,0,0}
 \definecolor{MAGENTA}{cmyk}{0,1,0,0}
 \definecolor{YELLOW}{cmyk}{0,0,1,0}
}

\makeatother

\usepackage{babel}
\begin{document}

\title{A Time-Asymmetric Process in Central Force Scatterings }

\author{Ramis Movassagh }

\email{ramis.mov@gmail.com}

\selectlanguage{english}%

\affiliation{Department of Mathematics, Northeastern University, Boston MA, 02115}

\date{\today}
\begin{abstract}
This article puts forth a process applicable to central force scatterings.
Under certain assumptions, we show that in attractive force fields
a high speed particle with a small mass speeding through space, statistically
loses energy by colliding softly with large masses that move slowly
and randomly. Furthermore, we show that the opposite holds in repulsive
force fields: the small particle statistically gains energy. This
effect is small and is mainly due to \textit{a}symmetric energy exchange
of the transverse (i.e., perpendicular) collisions. We derive a formula
that quantifies this effect (Eq.(\ref{eq:Delta_Efinal})). We then
put this work in a broader statistical context and discuss its consistency
with established results. 
\end{abstract}


\maketitle
\label{sec:Dynamical-Description}\textit{Dynamical Description-}
In addition to the well known gravitational and Coulomb, nearly all
other interactions in nature such as intermolecular forces and interaction
of vortices in superconductors are central \cite{LJones,blatter,ramis_roessler}.
In conservative fields, the central force on each particle can be
derived from a potential function by $F=-\mathbf{\mathbf{\nabla}}V\left(r\right)$
where $V(r)=\frac{\alpha}{r^{k}}$; $\alpha$ is the strength of the
interaction depending on the parameters of the problem, $k$ defines
the range of the interaction \cite{remark1}. $\alpha<0$ and $\alpha>0$
correspond to attractive and repulsive force fields respectively. 

In many applications, statistical inferences resulting from many-body
interaction is approximated by series of two-body scatterings \cite{chandra1,chandra2,chandra3,tyson}.
While there is an active frontier of numerical work on many-body simulations
\cite{Hut,Henon}, there are still interesting statistical inferences
that can be derived from close analysis of two-body collisions.

In this paper we study a fast small mass passing through a dilute
system of \textit{randomly moving} central forces, where changes in
the state of the small mass can be well approximated by a series of
two-body scatterings. We report on a net small effect in the statistical
change of the energy of the small mass (Eqs.(\ref{eq:DeltaE_v1},
\ref{eq:Delta_Efinal})). 

Consider a scattering, in the lab frame, between two interacting particles
$m_{1}$ and $m_{2}$ where $m_{2}$ is much lighter ($m_{2}\ll m_{1}$)
yet much faster ($v_{2}\gg v_{1}$) than $m_{1}$ but nevertheless
$m_{1}v_{1}^{2}\gg m_{2}v_{2}^{2}$. This in particular implies $m_{1}v_{1}\gg m_{2}v_{2}$.
For example, one can visualize a small comet ($m_{2}$) undergoing
a small angle scattering in the gravitational field of the planet
Jupiter ($m_{1}$). In a typical scattering $m_{1}$ is initially
moving. The questions we are interested in investigating are: What
statistically invariant features are shared by series of such scatterings
in randomly moving central force fields? Would many such small angle
scatterings have a net effect on the energy of $m_{2}$ under the
assumptions stated above? 

Statistical properties based on probabilities of encounters where
collisions are heads-on have been extensively studied as in the Fermi
acceleration mechanism \cite{fermi}. In contrast, the main contribution
to the effect herein is from \textit{transverse }collisions, where
the trajectory of the massive particle, roughly speaking, is perpendicular
to that of the small particle during the effective scattering. 

In the remainder of this section we heuristically describe the effect;
in the next section we analytically prove it and derive a formula.
Lastly, we discuss it in a larger statistical context. Throughout,
we refer to the Supplementary Material \cite{SM} for details when
needed. 

For the sake of concreteness take the potential to be attractive for
now. Let us consider two extreme cases that would convey the gist
of what underlies this work. In the first case the massive particle,
$m_{1}$, slowly and\textit{ }transversely veers away from the trajectory
of $m_{2}$ that is speeding by. In the second case, $m_{1}$ slowly
and transversely approaches the trajectory of $m_{2}$. 

\begin{figure}
\begin{centering}
\includegraphics[scale=0.27]{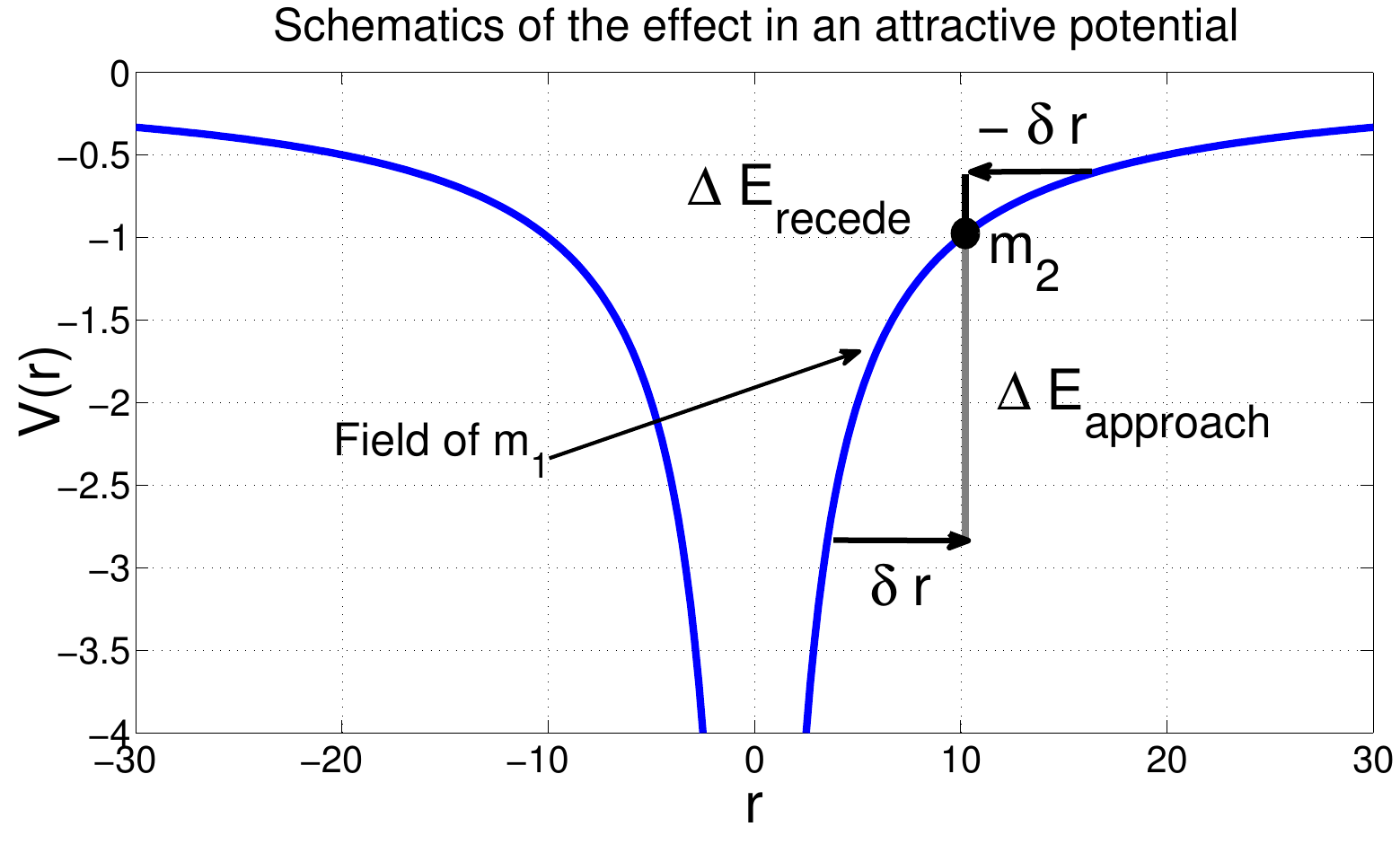}
\par\end{centering}

\caption{\label{fig:nonlinearity} $V(r)$ set up by $m_{1}$ shown in blue
with $\alpha<0$. The small mass $m_{2}$ is shown as a black circle.
When $m_{1}$ moves to the right (approaches $m_{2}$) by $\delta r$,
during the effective scattering, $m_{2}$ loses $\Delta E_{\mbox{approach}}$
of energy. When $m_{1}$ is moving to the left (away from $m_{2}$)
by the same amount $\delta r$, $m_{2}$ gains $\Delta E_{\mbox{recede}}$
of energy. Note: $|\Delta E_{\mbox{approach}}|>|\Delta E_{\mbox{recede}}|$.}
\end{figure}

In the first case where $m_{1}$ is moving away, $m_{2}$ falls into
the potential well of $m_{1}$ and so long as it is approaching the
point of minimum distance it gains kinetic energy. After passing this
point, $m_{2}$ starts climbing up the potential well and pays back
the gained kinetic energy by restoring it into the potential energy
of the two-body system. However, on the way out it climbs a potential
well that is effectively smaller than the one it fell into as $m_{1}$
is on average farther away from it ($\Delta E_{\mbox{recede}}$ in
Fig. \ref{fig:nonlinearity}). Therefore, in the case that the large
mass is \textit{transversely receding} away, the small particle emerges
with a \textit{gain} in the kinetic energy i.e., $\frac{1}{2}m_{2}|v_{m_{2}}|_{-\infty}^{2}<\frac{1}{2}m_{2}|v_{m_{2}}|_{+\infty}^{2}$. 

The exact opposite effect holds in the second case, where $m_{1}$
is moving towards $m_{2}$. In this case, $m_{2}$ enters the potential
well set up by $m_{1}$ and, as in the previous case, gains kinetic
energy so long as it is approaching the minimum distance between the
two masses. However, on the way out it faces a more demanding climb
as $m_{1}$ is on average closer to it and the potential well is steeper
and deeper than before ($\Delta E_{\mbox{approach}}$ in Fig. \ref{fig:nonlinearity}).
Therefore, in the case that the large mass is \textit{transversely
approaching}, the small particle emerges with a \textit{loss} in the
kinetic energy i.e. $\frac{1}{2}m_{2}|v_{m_{2}}|_{-\infty}^{2}>\frac{1}{2}m_{2}|v_{m_{2}}|_{+\infty}^{2}$
. 

The point however is that the two cases are not symmetric. \textit{The
decreasing of the magnitude of the force with distance breaks the
symmetry between the two cases}. This is shown in Fig. \ref{fig:nonlinearity}:
In an attractive force field, $m_{2}$ has a greater loss (in magnitude)
of energy when $m_{1}$ approaches it than a gain (in magnitude) when
$m_{1}$ recedes away from it. This asymmetry, deduced from dynamical
principles, has consequences for the statistical mechanics of $m_{2}$.

In repulsive force fields, $\alpha>0$, the potential in Fig. \ref{fig:nonlinearity}
flips about the horizontal axis. Therefore the phenomenology is the
exact opposite. Namely, the particle has a larger gain than loss. 

So far we have described a purely dynamical phenomena where $m_{2}$
collides softly and transversely with $m_{1}$ where a very fast small
particle (e.g., an electron) zips through a dilute soup of big masses
(e.g, massive ions or stars in a galaxy) that randomly either approach
it or move away from it. The small mass statistically loses (gains)
energy to (from) the big masses when the force fields are attractive
(repulsive). 

We are considering an standard elastic collision \cite{L_Mechanics}.
Let $\mathbf{v_{1}}$ and $\mathbf{\mathbf{\mathbf{v}}_{2}}$ be the
velocities of $m_{1}$ and $m_{2}$ respectively in the lab frame
and let $\mathbf{V}=\mathbf{v_{2}}-\mathbf{\mathbf{v}_{1}}$. Denote
by $\mathbf{n}_{0}^{+}$ the unit vector in the direction of the velocity
of $m_{2}$ in the center of mass after the collision which is parallel
to $\mathbf{V}$. Then the velocities of the two particles after the
collision (distinguished by primes) are 

\begin{eqnarray}
\mathbf{v_{1}'} & = & -m_{2}V\mathbf{\mathbf{n}_{0}^{+}}/(m_{1}+m_{2})+\mathbf{V_{g}},\label{eq:v1lab}\\
\mathbf{v_{2}'} & = & m_{1}V\mathbf{n}_{0}^{+}/(m_{1}+m_{2})+\mathbf{V_{g}},\label{eq:v2lab}
\end{eqnarray}
where $\mathbf{V_{g}}=\left(m_{1}\mathbf{v_{1}}+m_{2}\mathbf{v_{2}}\right)/\left(m_{1}+m_{2}\right)\approx\mathbf{v_{1}}$
is the velocity of the center of mass. $ $No further information
about the collision can be obtained from the laws of conservation
of momentum and energy. The direction of the unit vector $\mathbf{n}_{0}^{+}$
depends on the particular law of interaction and positions during
the collision. 

We assume the massive particles are far enough from one another that
a sequence of two-body scatterings would be an adequate approximation
\cite{chandra4}. Let us denote by $\Delta E$ the change in the energy
of $m_{2}$ before and after any given collision $\Delta E=\frac{m_{2}}{2}\left(v_{2}^{'2}-v_{2}^{2}\right)$,
which is positive when $m_{2}$ gains energy in collision and is negative
when it loses energy. Using Eq.(\ref{eq:v2lab}) we find 

\begin{equation}
\Delta E=\mu V\mathbf{V_{g}.}\left[\mathbf{n}_{0}^{+}-\mathbf{n}_{0}^{-}\right]\approx\mu V\mathbf{v_{1}.}\mathbf{n}\label{eq:delKE}
\end{equation}
where, $\mathbf{n}\equiv\mathbf{n}_{0}^{+}-\mathbf{n}_{0}^{-}$, $\mu=\frac{m_{1}m_{2}}{m_{1}+m_{2}}\approx m_{2}$
is the reduced mass, $\mathbf{n}_{0}^{-}$ and $\mathbf{n}_{0}^{+}$
denote the unit vectors pointing in the direction of motion of $m_{2}$
before and after the collision in the center of mass. Let us, once
again, look at the two special cases discussed above. First consider
an attractive force field. Suppose $\mathbf{v_{1}}$ lies on the line
of the minimum distance as shown in Fig. \ref{fig:General_scattering(a,b)}a.
Clearly if $m_{1}$ recedes away from $m_{2}$ then $\mathbf{v_{1}}$
points in the same direction as $\mathbf{n}$ and the dot product
on the right hand side of Eq.(\ref{eq:delKE}) is positive, whereas
if $\mathbf{v_{1}}$ and $\mathbf{n}$ point in opposite directions
the right hand side is negative. In the case of repulsion the signs
would be the opposite (see Fig. \ref{fig:General_scattering(a,b)}b).

\begin{figure}
\begin{raggedright}
\includegraphics[scale=0.16]{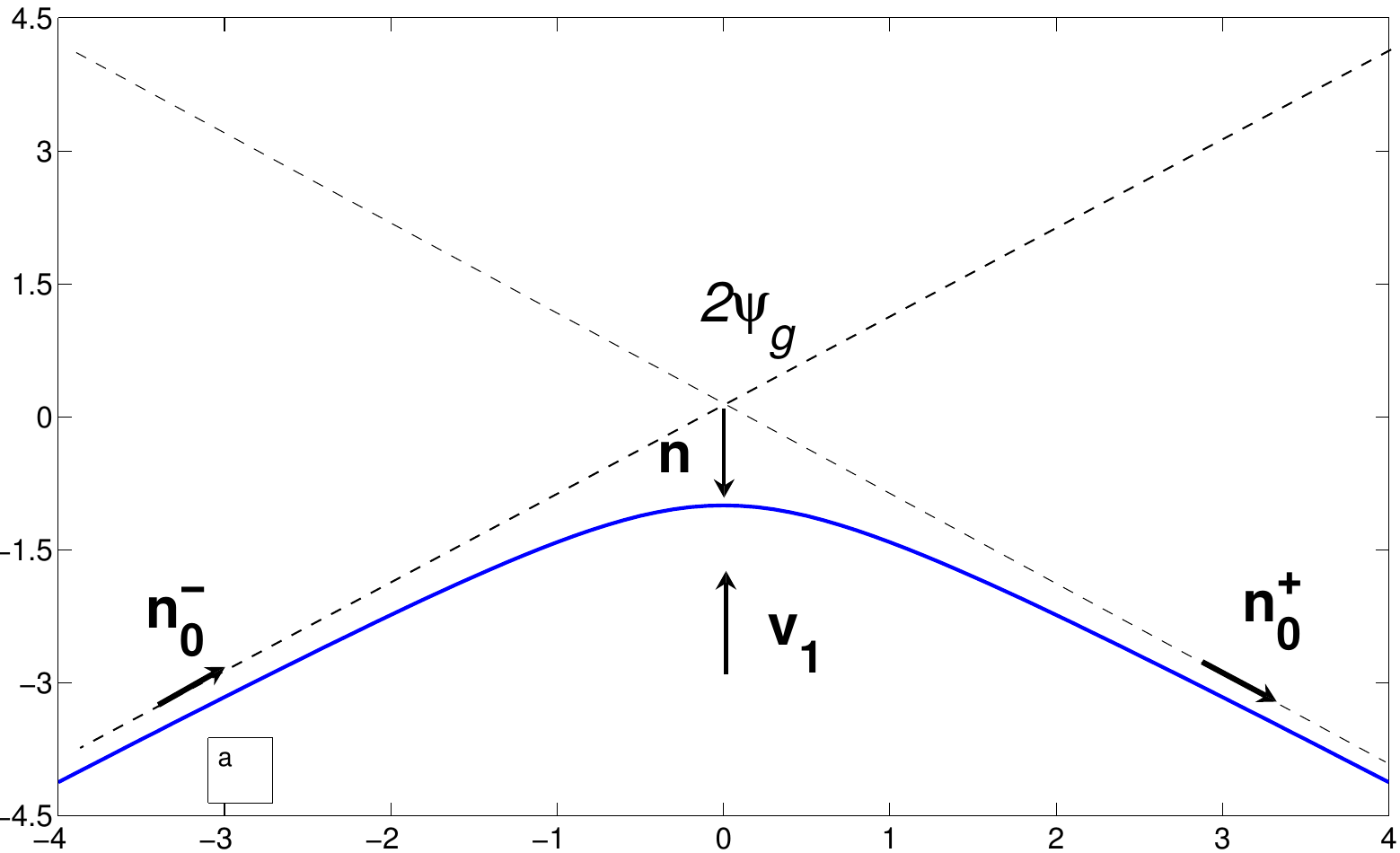}\includegraphics[scale=0.16]{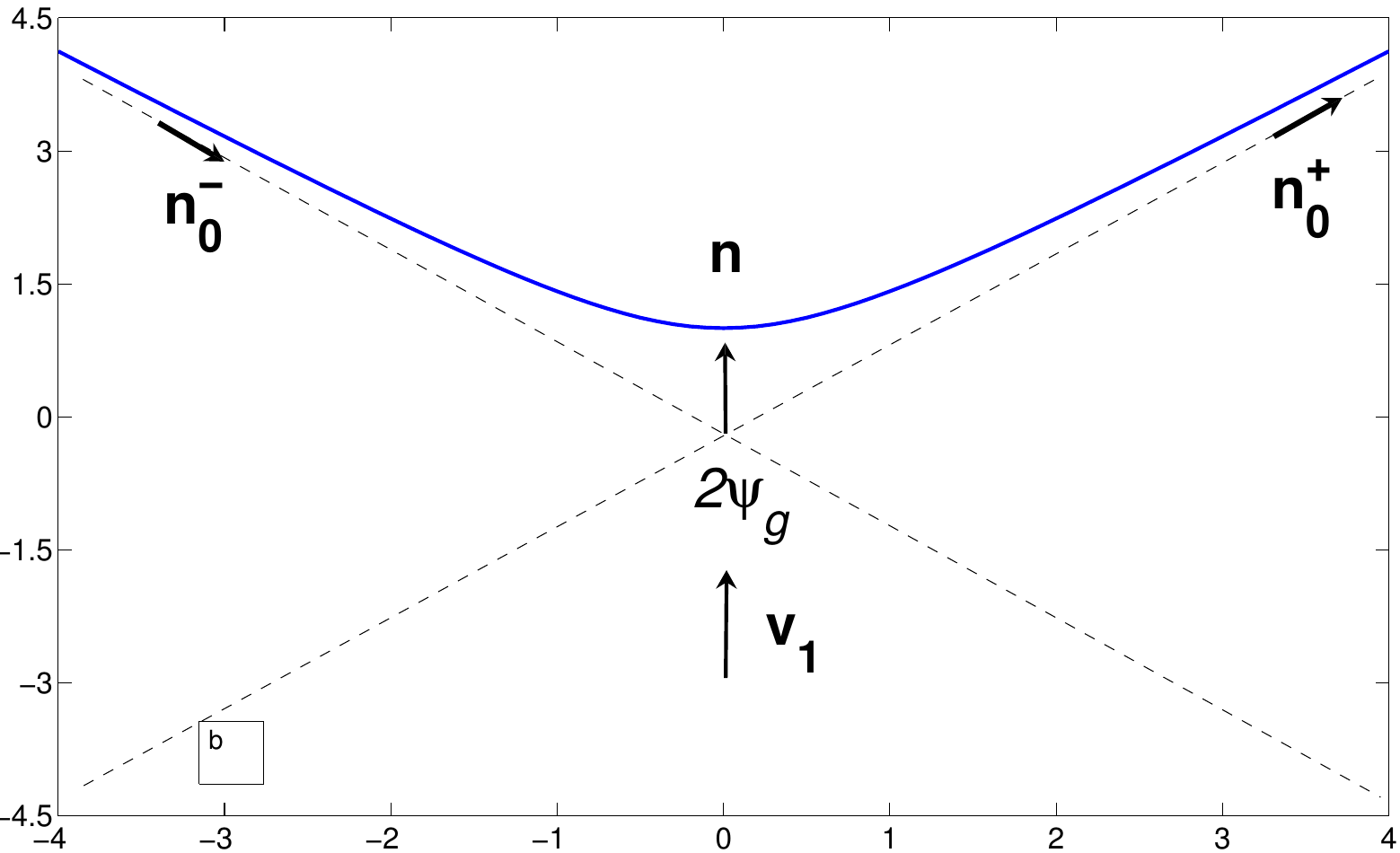}
\par\end{raggedright}

\caption{\label{fig:General_scattering(a,b)}The relationship between the vectors
in Eq.(\ref{eq:delKE}) a) an attractive force field and b) repulsive
force field. }
\end{figure}

What we now argue is that the two cases are not symmetric. That is
the kinetic energy loss (gain) in the approaching case is larger than
the gain (loss) in the receding case for an(a) attractive (repulsive)
potential. From Figs. (\ref{fig:General_scattering(a,b)}a-b) we see
that $\left|\mathbf{n}\right|=\textrm{2cos}\psi_{g}$. In the attractive
case (Fig. \ref{fig:General_scattering(a,b)}a), if $m_{1}$ moves
towards $m_{2}$ the angle between the asymptotes, $2\psi_{g}$, is
smaller than it would be if $m_{1}$ moved away as the minimum distance
is smaller. Therefore, $n=2\cos\psi_{g}$ is larger in the approaching
case. For very high speed encounters, $\mathbf{\left|v_{2}\right|}\gg\mathbf{\left|v_{1}\right|}$,
we can approximate $v$ to be the same in the two cases (Figs. \ref{fig:General_scattering(a,b)}
a and b), therefore Eq.(\ref{eq:delKE}) becomes:

\begin{equation}
\Delta E=\mu V\mathbf{V_{g}.}\mathbf{n\simeq\left\{ \begin{array}{c}
\mathit{-2\mu Vv_{1}}\left|\textrm{cos}\psi_{g}^{a}\right|\\
\hphantom{-}\mathit{2\mu Vv_{1}}\left|\textrm{cos}\psi_{g}^{r}\right|
\end{array}\right.},\label{eq:asymmetry}
\end{equation}
with $\left|\textrm{cos}\psi_{g}^{a}\right|>\left|\textrm{cos}\psi_{g}^{r}\right|$
(see following Eq.(\ref{eq:DelE_recede_firstorder})). 

\textit{Analytical Derivation-} The heuristic arguments above apply
to general central forces. For a two-body scattering based on laws
of conservation of linear momentum and energy applicable to any law
of interaction one can show \cite{Gryzinski1965_I,Gryzinski1965_II}
\begin{figure}[h]
\centering{}\includegraphics[scale=0.28]{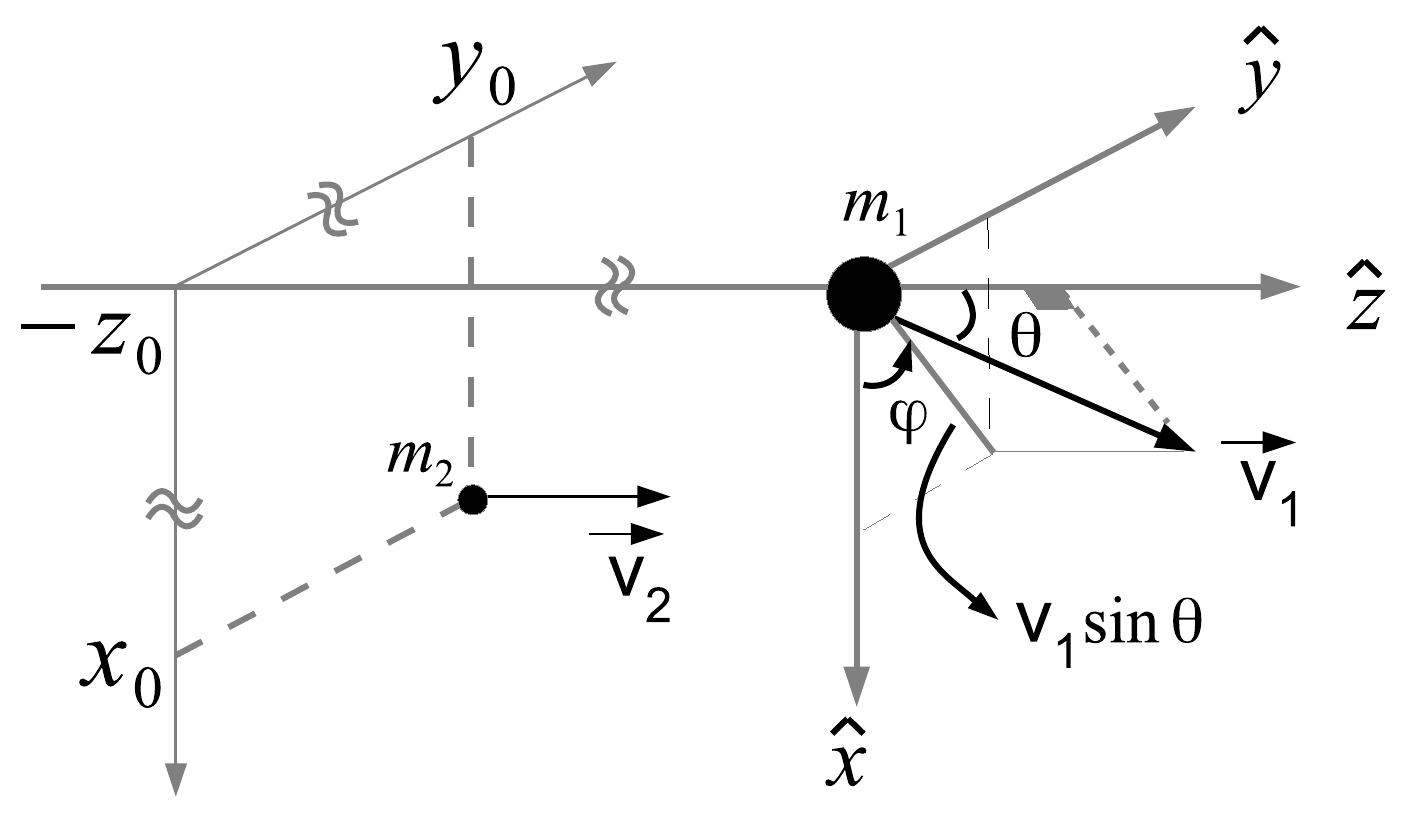}\caption{\label{fig:Initial-configuration.}Initial configuration.}
\end{figure}

\begin{eqnarray}
\Delta E & = & 2a\sin\psi_{g}\cos\psi_{g}\cos\Theta-b\cos^{2}\psi_{g}\label{eq:DelE_raw}\\
a & \equiv & \mu v_{1}v_{2}\sin\theta,\nonumber \\
b & \equiv & K_{12}\left[m_{2}v_{2}^{2}-m_{1}v_{1}^{2}+\left(m_{1}-m_{2}\right)v_{1}v_{2}\cos\theta\right],\nonumber \\
K_{12} & \equiv & 2\left(m_{1}m_{2}\right)/\left(m_{1}+m_{2}\right)^{2}.\nonumber 
\end{eqnarray}
where $\Theta$ is the angle between the orbital and fundamental planes.
The orbital plane is the plane perpendicular to the angular momentum
vector and the fundamental plane contains $\mathbf{v_{1}}$ and $\mathbf{v_{2}}$.
Eq.(\ref{eq:DelE_raw}) is general and the law of interaction only
enters through $\psi_{g}$. 

What is needed is the full formulation of the scattering in the lab
frame. For any given encounter, let us set up a coordinate system
with the $\boldsymbol{\hat{z}}$ pointing in the direction of $\mathbf{v_{2}}$
and put $m_{1}$ at its origin as shown in  Fig. \ref{fig:Initial-configuration.}.
The initial state of $m_{2}$ is $\boldsymbol{r_{2}}=\left(x_{0},y_{0},-z_{0}\right)$
and $\mathbf{v_{2}}=\left(0,0,v_{2}\right)$ where we explicitly took
$z_{0}>0$. The initial state of $m_{1}$ is $\boldsymbol{r_{1}}=\boldsymbol{0}$
$ $ and $\mathbf{v_{1}}=v_{1}\left(\sin\theta\cos\varphi,\sin\theta\sin\varphi,\cos\theta\right)$;
consequently $V=\left\{ v_{1}^{2}+v_{2}^{2}-2v_{1}v_{2}\cos\theta\right\} ^{1/2}.$
The impact parameter $D$ is the minimum distance between the masses
in the absence of interaction ($\alpha=0$). Both $D$ and $\Theta$
are functions of the relative states of the two masses \cite{SM},

\begin{widetext}

\begin{eqnarray}
D & = & \left\{ D_{0}^{2}-\frac{v_{1}\sin\theta}{V^{2}}\left[\frac{v_{1}}{2}\sin\theta\left(D_{0}^{2}-2z_{0}^{2}+\left(x_{0}^{2}-y_{0}^{2}\right)\cos2\varphi+2x_{0}y_{0}\sin2\varphi\right)+2v_{2}z_{0}\left(1-\frac{v_{1}}{v_{2}}\cos\theta\right)r_{0}\left(\varphi\right)\right]\right\} ^{1/2},\label{eq:D}\\
\cos\Theta & = & \frac{r_{0}\left(\varphi\right)\left(v_{2}-v_{1}\cos\theta\right)-v_{1}z_{0}\sin\theta}{B},\label{eq:cosTheta}\\
B & \equiv & \left\{ \left[v_{2}x_{0}-v_{1}\left(x_{0}\cos\theta+z_{0}\sin\theta\cos\varphi\right)\right]^{2}+v_{1}^{2}\sin^{2}\theta\left(-x_{0}\sin\varphi+y_{0}\cos\varphi\right)^{2}\right.\\
 & + & \left.\left[y_{0}\left(v_{2}-v_{1}\cos\theta\right)-v_{1}z_{0}\sin\theta\sin\varphi\right]^{2}\right\} ^{1/2}.\nonumber 
\end{eqnarray}
\end{widetext}where $D_{0}^{2}\equiv x_{0}^{2}+y_{0}^{2}$ and $r_{0}\left(\varphi\right)\equiv x_{0}\cos\varphi+y_{0}\sin\varphi$.
From now on we confine to gravitational and Coulomb interactions \cite{goldstein}.
The strength of interaction, $\alpha$, for gravitational and Coulomb
interactions are $Gm_{1}m_{2}$ and $\frac{q_{1}q_{2}}{4\pi\varepsilon_{0}}$
respectively and $\cos\psi_{g}=\frac{\alpha}{\mu V^{2}D}/\sqrt{1+\left(\frac{\alpha}{\mu V^{2}D}\right)^{2}}.$

The orbital and fundamental planes coincide for $y_{0}=0$, $\theta=\frac{\pi}{2}$
and $\varphi=0,\pi$. The special cases are: $\cos\Theta=1$ for $\varphi=0$
(approach) and $\cos\Theta=-1$ for $\varphi=\pi$ (recede). Ignoring
terms of $\mathcal{O}\left(\frac{v_{1}}{v_{2}}\right)^{2}$ and higher,
we find 

\begin{eqnarray}
\Delta E_{approach} & = & \frac{2\alpha}{D_{0}}\frac{v_{1}}{v_{2}}+\mathcal{O}\left(\frac{v_{1}}{v_{2}}\right)^{2},\label{eq:DelE_appr_firstorder}\\
\Delta E_{recede} & = & -\frac{2\alpha}{D_{0}}\frac{v_{1}}{v_{2}}+\mathcal{O}\left(\frac{v_{1}}{v_{2}}\right)^{2}.\label{eq:DelE_recede_firstorder}
\end{eqnarray}
Further, we find $\cos\psi_{g}^{a}=\frac{\alpha}{\mu D_{0}v_{2}^{2}}\left\{ 1+\left(\frac{v_{1}}{v_{2}}\right)\left[\frac{z_{0}}{D_{0}}\left(\frac{x_{0}}{D_{0}}\right)\right]\right\} $
and $\cos\psi_{g}^{r}=\frac{\alpha}{\mu D_{0}v_{2}^{2}}\left\{ 1-\left(\frac{v_{1}}{v_{2}}\right)\left[\frac{z_{0}}{D_{0}}\left(\frac{x_{0}}{D_{0}}\right)\right]\right\} $.
Some conclusions can be drawn to first order in $\frac{v_{1}}{v_{2}}$: 

1. For $\alpha<0$ the transverse collisions lead to small yet nonzero
energy gain (loss) for the receding (approaching) collisions. For
$\alpha>0$ the transverse collisions lead to small yet nonzero energy
gain (loss) for the approaching (receding) collisions. This proves
the heuristic arguments we gave earlier (depicted in Fig.\ref{fig:nonlinearity}).

2. To investigate any asymmetry in $\Delta E$ (Fig. \ref{fig:nonlinearity})
one needs to go beyond $\mathcal{O}\left(\frac{v_{1}}{v_{2}}\right)$;
see below. 

3. We see that $\cos\psi_{g}^{r}<\cos\psi_{g}^{a}$ as argued above
in and following Eq.(\ref{eq:asymmetry}). 

We now calculate the \textit{ensemble} average $\langle\Delta E\rangle_{\mathbf{\mathbf{v}_{1}}}$,
where the subscript indicates the random variable with respect to
which we average. It is useful to call $m_{2}$ the \textit{test particle}
undergoing small angle scattering through many collisions with \textit{field
particles} of mass $m_{1}$. Let $f\left(\theta,\varphi\right)d\theta d\varphi$
denote the probability that the velocity vector of a field particle
has direction determined by angles $\theta$ and $\varphi$. For isotropically
moving field particles $f\left(\theta,\varphi\right)=\frac{1}{2}\sin\theta\left(\frac{1}{2\pi}\right)$.
Calculating $\Delta E$ as given by Eq.(\ref{eq:DelE_raw}) and using
Eqs.(\ref{eq:D},\ref{eq:cosTheta}) and ignoring term of order $\mathcal{O}\left(\frac{v_{1}}{v_{2}}\right)^{3}$
and higher \cite{SM}, we find $\langle\Delta E\rangle_{\theta,\varphi}\approx\frac{2}{3}\frac{\alpha}{D_{0}}\frac{z_{0}}{D_{0}}\left(\frac{v_{1}}{v_{2}}\right)^{2}+\mathcal{O}\left(\frac{v_{1}}{v_{2}}\right)^{3}$
.

It is found under rather general conditions that the speed of the
field particles is Gaussian distributed $\frac{4j^{3}}{\sqrt{\pi}}e^{-j^{2}v_{1}^{2}}v_{1}^{2}dv_{1},$
where $j$ is related to the variance $\sigma$ by $j^{2}\equiv\frac{1}{\sqrt{2}\sigma^{2}}$
\cite[Eq: 2.353]{chandra4}. The integration with respect to $\rho\left(v_{1},\theta,\varphi\right)=\frac{j^{3}}{\pi^{3/2}}\sin\theta e^{-j^{2}v_{1}^{2}}v_{1}^{2}dv_{1}d\theta d\varphi$
gives the ensemble average,

\begin{equation}
\langle\Delta E\rangle_{\mathbf{\mathbf{v}_{1}}}=\frac{2\alpha}{D_{0}j^{2}v_{2}^{2}}\frac{z_{0}}{D_{0}}.\label{eq:DeltaE_v1}
\end{equation}
For a density of field particles $n$, adopting a cylindrical coordinate
system in which the radius is $D_{0}$ and $\phi$ is the azimuthal
angle, the number of field particles passing $m_{2}$ per unit time
is $n\int d^{3}x/dt=n\left(\frac{dz}{dt}\right)\int dA$, which to
leading order is $nv_{2}\int d\phi\int D_{0}dD_{0}$. Performing the
integral we find

\[
nv_{2}\int d\phi\int D_{0}dD_{0}\langle\Delta E\rangle_{\mathbf{\mathbf{v}_{1}}}=\frac{4\pi n\alpha z_{0}}{j^{2}v_{2}}\int\frac{dD_{0}}{D_{0}}
\]

The integral diverges for $D_{0}\rightarrow\infty$; this is natural
as the force is long range and by definition very distant encounters
need to be taken into account. It also diverges for $D_{0}\rightarrow0$,
which violates the distant encounter assumptions $\frac{\alpha}{\mu DV^{2}}\ll1$.
However, in a given problem, there is a natural $D_{min}$ that ensures
small angle scattering and a $D_{max}$, depending on the density
of field particles. Further a factor of $2$ or $3$ error in choosing
$D_{max}/D_{min}$ does not affect the calculation of relaxation times
by much \cite{Divergent_Integral}.

A typical time scale between collisions is $\delta t\sim2z_{0}/v_{2}\sim n^{-1/3}/v_{2}$.
The statistical change in energy of $m_{2}$ after $N$ encounters,
denoted by $\langle\Delta E\rangle$ is $\frac{4\pi n\alpha z_{0}}{j^{2}v_{2}}\ln\left(\frac{D_{max}}{D_{min}}\right)N\delta t$
giving the desired result

\begin{equation}
\langle\Delta E\rangle\approx\frac{2\pi\alpha Nn^{1/3}}{j^{2}v_{2}^{2}}\ln\left(\frac{D_{max}}{D_{min}}\right).\label{eq:Delta_Efinal}
\end{equation}
Note that attractive interactions, i.e., $\alpha<0$, yield an average
loss and $\alpha>0$ an average gain. 

\textit{Statistical Context-} A phenomena worth considering is dynamical
friction \cite{chandra1,chandra2}. Dynamical friction, however, is
like Brownian motion \cite{einstein} as a big mass enters a medium
of many smaller particles and slows down as a result. But it is distinct
from Brownian motion as the interactions are long ranged. It is found
that in dynamical friction ``only stars with velocities less than
the one under consideration contribute to the effect'' \cite{chandra3}
and \cite[p. 299]{chandra4}. 

The main requirement here is that the test particle scatters from
the\textit{ time-dependent }field\textit{ }of, in comparison massive,
particles that are moving randomly and slowly in space, through a
series of small-angle scatterings. 

At first sight this effect seems to violate the equipartition of energy
because a low energy particle ``heats up'' the medium of much larger
particles that have higher energies. We are working with a non-equilibrium
process in an open system. In attractive potentials the small particle
starts from non-equilibrium initial conditions and through a series
of scatterings it statistically loses energy till a final scattering
where the energy in the center of mass is negative. There on the small
particle would have a bounded orbit about that final scatterer. This
corresponds to the breakdown of small angle scattering assumption
we have made. In plasma physics this is known as shielding and in
astrophysics it corresponds to capturing of a comet by a center of
force. In repulsive potentials the energy of the small particle grows
till the relativistic effects become significant and the transfer
of energy between the small particle and the scatterers becomes of
order unity with respect to the initial energy \cite[section 13]{LandauLifshitzFields}.
Hence the effect is not an equilibration process and is applicable
to systems where the assumptions of small angle scattering, as well
as, $m_{1}\ll m_{2}$, $v_{1}\gg v_{2}$ but nevertheless $m_{1}v_{1}^{2}\ll m_{2}v_{2}^{2}$
hold. The small angle scattering assumption is bound to break on time
scales comparable to the relaxation time to equilibrium. 

If we relax the assumption $m_{1}\gg m_{2}$ and let $m_{1}$ be comparable
to $m_{2}$, then for small angle scatterings and $v_{2}\gg v_{1}$
one finds that $\langle\Delta E\rangle<0$ regardless of the sign
of $\alpha$ as expected (\cite[Section 3.1]{SM}). 

It would be interesting to analyze the effect of micro-dynamics of
the structure in the universe on the frequency shift of photons coming
from distant sources. We expect a small loss of energy for photons
that undergo dynamical weak lensing \cite{weinberg,tyson}. We hope
this work helps better understanding of rapid structure formation
\cite{peebles}, high energy cosmic rays \cite{cosmicRay} and redshift
problem \cite{Reyes}.
\begin{acknowledgments}
I thank Richard V. Lovelace, Peter W. Shor, Jack Wisdom and especially
Jeffrey Goldstone for discussions. I also thank the editor, the anonymous
referee, Mehran Kardar, Frank Wilczek, Otto E. Rossler, and Eduardo
Cuervo-Reyes. This work was partly supported by the National Science
Foundation through grant number CCF-0829421.\end{acknowledgments}

\section*{Supplementary Material. Two-Particle Collisions: The Laboratory Frame
Formulation}

Let us consider an encounter between two particles of masses $m_{1}$
and $m_{2}$. If the initial conditions are specified, the dynamics
in principle is fully determined. To fully specify the state at any
time one needs $12$ parameters; $3$ positions and $3$ momenta per
particle. A system of $n$ interacting particles that are isolated
otherwise have conservation of energy, linear and angular momentum
intact regardless of the particular laws of interaction among the
constituents. This applies to the two body problem ($n=2$) as well
and provides us with $7$ constraints, $1$ energy $3$ linear and
$3$ angular momenta. Therefore there are $5$ other degrees of freedom
that depend on the geometry of the encounter and the law of interaction.
\begin{figure}
\centering{}\includegraphics[scale=0.9]{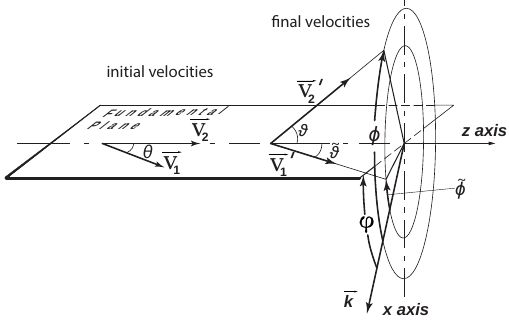}\caption{Initial and final velocities with respect to the laboratory system.
$\boldsymbol{v_{1}}$ and $\boldsymbol{v_{2}}$ define the fundamental
plane.}
\end{figure}

Below we denote vectors in boldface (e.g., $\boldsymbol{v_{1}}$ is
a vector with magnitude $v_{1}$). Let the velocities before the collision
be $\boldsymbol{v_{1}}$ and $\boldsymbol{v_{2}}$ and after $\boldsymbol{v_{1}'}$
and $\boldsymbol{v_{2}'}$. We take the $z$ axis of the laboratory
system to coincide with the initial direction of particle $2$, which
we call the {\it test particle}. As a result of interaction with
particle $1$, which we call the {\it field particle}, the velocity
of particle $2$ changes in magnitude and direction. We denote the
change in the energy of particle $2$ by $\Delta E$, the direction
of velocity after the scattering by $\vartheta$ and $\phi$ \cite[Figure 1]{Gryzinski1965_I}.
Similarly we assign $\Delta\tilde{E}$, $\tilde{\vartheta}$ and $\tilde{\phi}$
for the field particle. It is clear that the result of collision depends
on the laws of interaction as well as the geometry of encounter. To
describe the geometry of the encounter, we need four geometrical quantities
(one linear quantity and three angular ones \cite[Figure 2]{Gryzinski1965_I})

1. the collision impact parameter $D$

2. the angle $\theta$ which is the angle between $\boldsymbol{v_{1}}$
and $\boldsymbol{v_{2}}$; the plane containing these two vectors
is called the \textit{fundamental plane}

3. the angle $\Theta$, which is the angle formed by the segment $D$
with the fundamental plane. Equivalently this is the angle between
the \textit{orbital plane} and the fundamental plane. The orbital
plane is the plane perpendicular to the angular momentum vector and
contains the relative velocity before and after the collision.

4. the angle $\varphi$, which is the angle describing the position
of the fundamental plane with respect to rotation about the $z$ axis. 

\begin{figure}[H]
\centering{}\includegraphics[scale=0.9]{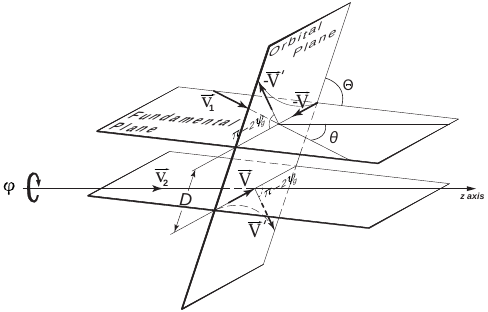}\caption{Space diagram of the encounter. The orbital plane contains $\boldsymbol{V}$
and $\boldsymbol{V'}$. The law of interaction is completely encoded
in $\psi_{g}$.}
\end{figure}

The range of these parameters are \cite[Figure 2]{Gryzinski1965_I}

\begin{eqnarray*}
0 & \le D\le & \infty,\\
0 & \le\theta\le & \pi,\\
0 & \le\Theta\le & 2\pi,\\
0 & \le\varphi\le & 2\pi.
\end{eqnarray*}
Now we shall derive the dependence of $\Delta E$, $\vartheta$,$\phi$
and $\Delta\tilde{E}$, $\tilde{\vartheta}$ ,$\tilde{\phi}$ in term
of the geometrical quantities. The fifth parameter, which encodes
the dependence on the law of interaction is $\psi_{g}$-- the scattering
angle in the center of mass frame.

\subsection{Law of interaction and $\psi_{g}$}

\begin{figure}
\begin{centering}
\includegraphics[scale=0.26]{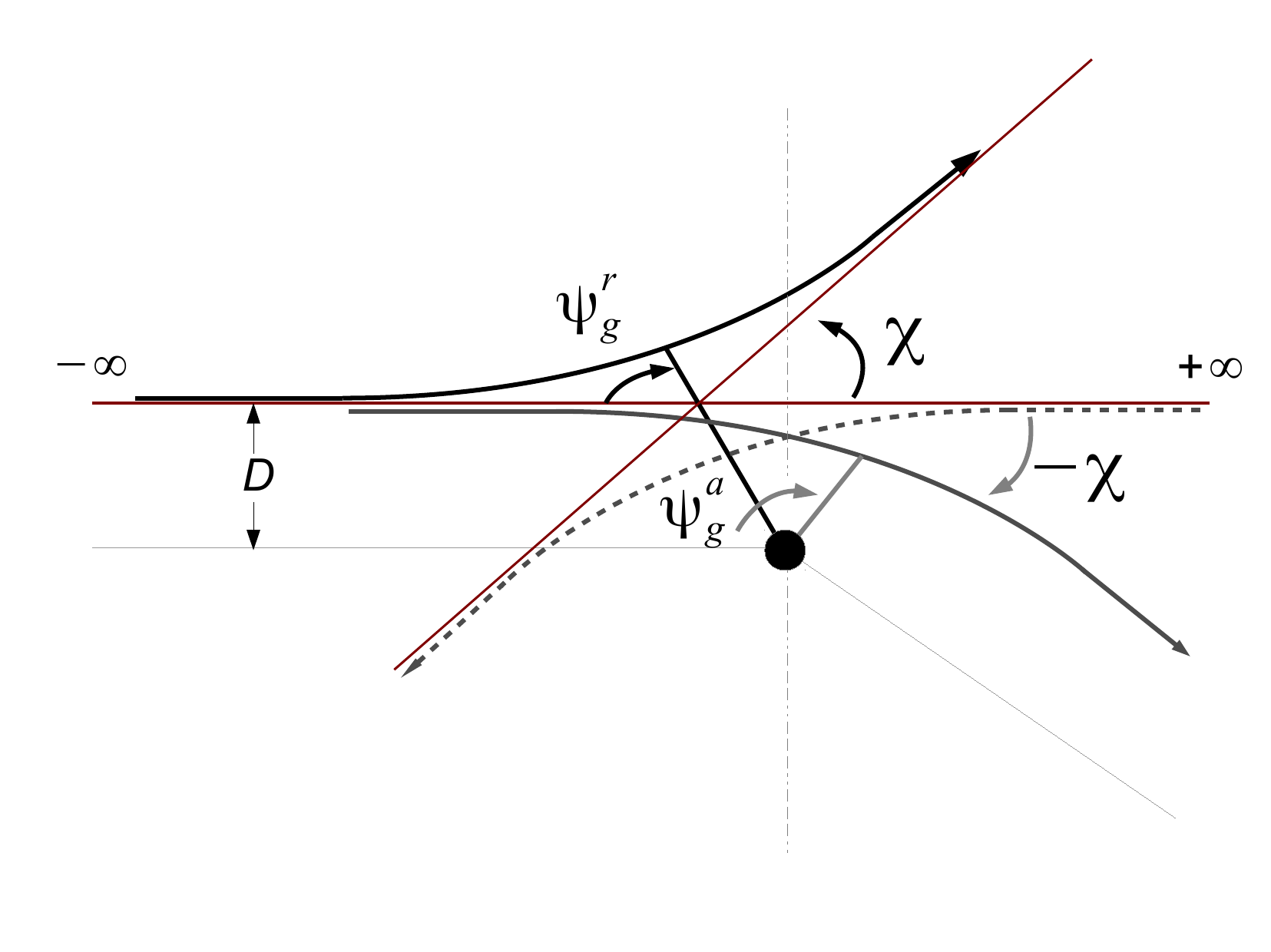}
\par\end{centering}

\centering{}\caption{\label{fig:Scattering-and-deflection}Scattering and deflection angles
shown in $\psi_{g}$ and $\chi$ respectively.}
\end{figure}

The dependence on the interaction law only enters through the trigonometric
functions of the angle $\psi_{g}$, i.e., the angle describing the
scattering of the \textit{reduced mass} $\mu$ in the center of mass
system. Therefore, for a unified formulation that applies to repulsive
and attractive force laws given by the potential $U\left(r\right)=\frac{\alpha}{r^{k}}$
where 
\begin{eqnarray*}
\alpha & < & 0\mbox{ }:\quad\mbox{attractive}\\
\alpha & > & 0\mbox{ }:\quad\mbox{repulsive}
\end{eqnarray*}
it is necessary that the sign of the interaction enters the formulas
derived for $\psi_{g}$. The importance of this is especially pronounced
when considering collision statistics. In calculating a scalar quantity
$\psi_{g}$ and ignoring the origin with respect to which it extends
ignores the sign of the interaction (see Figure \ref{fig:Scattering-and-deflection}).

The scattering for $k=1$ corresponds to hyperbolic trajectories in
the center of mass frame. For the sake of concreteness let us consider
a Coulomb scattering where we have a force field with charge $+Ze$
pinned down, shown as a black center in the Figure \ref{fig:Scattering-and-deflection}.
Take that be a focal point of the hyperbola which has two branches
one shown in black and the other in grey dashed curve. We put an arrow
head to emphasize the temporal order of the points on each branch.
The solid black branch corresponds to a positive particle, say a positron
$+e$, coming from $-\infty$ (i.e., left) and scattering off. The
grey dashed branch corresponds to an electron coming from $+\infty$
(i.e., right). The minimum distances from the center of the force
in both cases lie on the same line, albeit the magnitudes are different
($-e$ gets closer to the center of the force).

The black branch describes the scattering in a repulsive force field.
How does the scattering look if we consider an electron coming from
$-\infty$? This corresponds to the reflection of the dashed grey
branch, shown as a solid grey, about the vertical axis piercing the
center of the force (the vertical axis is shown by a thin dashed line).
The geometry makes it clear that $\psi_{g}$ for this case is greater
than $\frac{\pi}{2}$, whereas it is less than $\frac{\pi}{2}$ for
the repulsive force field and it is exactly $\frac{\pi}{2}$ when
there is no interaction. The calculation of the angle correspondingly
gives a negative value for $\cos\psi_{g}$ in an attractive force
field and a positive value in a repulsive force field. 

We are concerned with interactions that are central, they depend only
on the distance between the particles, then the relation describing
the angle $\psi_{g}$ is relatively simple

\begin{equation}
\psi_{g}=\intop_{r_{min}}^{\infty}\frac{\left(D/r^{2}\right)dr}{\left\{ 1-\frac{2U\left(r\right)}{\mu V^{2}}-\left(D/r\right)^{2}\right\} ^{1/2}},\label{eq:psi_g_general}
\end{equation}
where $\mu=\frac{m_{1}m_{2}}{m_{1}+m_{2}}$, $V$and $D$ have the
same meaning as above and $U\left(r\right)$ is the potential function
of the two particles and $r_{min}$ is the distance of closest approach
of $\mu$ to the center of force determined by the turning point at
which $\dot{r}=0$. Let $\rho=\frac{D}{r}$, the potential becomes
$\tilde{U}\left(\rho\right)=\frac{\alpha\rho^{k}}{D^{k}}$ and the
scattering angle 

\begin{equation}
\psi_{g}=\intop_{0}^{\rho_{max}}\frac{d\rho}{\left\{ 1-\frac{2\alpha}{\mu V^{2}}\left(\frac{\rho}{D}\right)^{k}-\rho^{2}\right\} ^{1/2}},\label{eq:psi_2}
\end{equation}
where $\rho_{max}$ is the positive root of the quantity in the denominator.
In scattering problems energy in the center of mass $E>0$ hence $\frac{\mu}{2}V^{2}+U>0$
which implies $-\frac{2U}{\mu V^{2}}<1.$ For $U\left(r\right)=\frac{\alpha}{r^{k}}$
this becomes $-\frac{2\alpha}{\mu V^{2}r^{k}}<1$. The inequality
is trivially satisfied for $\alpha>0$ and for attractive potentials
implies $0\le\frac{2\left|\alpha\right|}{\mu V^{2}r^{k}}<1$ and $2\left|\alpha\right|<\mu V^{2}r^{k}$.
Eq. \ref{eq:psi_2} for $k=1$ is an elementary integral of form

\begin{eqnarray*}
\int\frac{dx}{\sqrt{a+bx+cx^{2}}} & = & \frac{1}{\sqrt{-c}}\cos^{-1}\left(-\frac{b+2cx}{\sqrt{\lambda}}\right)\\
\lambda & \equiv & b^{2}-4ac,
\end{eqnarray*}
substituting we find

\begin{eqnarray*}
\psi_{g} & = & \cos^{-1}\frac{\epsilon+\rho}{\sqrt{1+\epsilon^{2}}}\\
\epsilon & \equiv & \frac{\alpha}{\mu V^{2}D}
\end{eqnarray*}

Further

\[
\rho_{max}=-\epsilon+\sqrt{1+\epsilon^{2}}.
\]
Performing the integral we get

\begin{eqnarray}
\cos\psi_{g} & = & \frac{\epsilon}{\sqrt{1+\epsilon^{2}}}=\frac{\frac{\alpha}{\mu V^{2}D}}{\sqrt{1+\left(\frac{\alpha}{\mu V^{2}D}\right)^{2}}}\label{eq:cosPsi_k=00003D1}\\
\sin\psi_{g} & = & \frac{1}{\sqrt{1+\epsilon^{2}}}=\frac{1}{\sqrt{1+\left(\frac{\alpha}{\mu V^{2}D}\right)^{2}}}\label{eq:sinPsi_k=00003D1}
\end{eqnarray}

Below we mainly deal with the case of $k=1$. Though it is irrelevant
for most of what follows, as an aside, for $k=2$, the integration
yields 

\begin{equation}
\psi_{g}=\frac{\pi}{2}\frac{1}{\left\{ 1+\left(\frac{2\alpha}{\mu V^{2}D^{2}}\right)\right\} ^{1/2}}.\label{eq:Psi_k=00003D2}
\end{equation}
Moreover, we mention that in the case of a collision of two impenetrable
spheres of radii $R_{1}$ and $R_{2}$, Eq. \ref{eq:psi_2} gives

\begin{eqnarray*}
\sin\psi_{g} & = & 2D/\left(R_{1}+R_{2}\right),\\
\cos\psi_{g} & = & \left\{ 1-\left[2D/\left(R_{1}+R_{2}\right)\right]^{2}\right\} ^{1/2}.
\end{eqnarray*}
\vspace{0.4cm}

\section*{Dynamics of a Two-Particle Encounter}

The relations we derive in this section are in the most general form
based on laws of conservation of linear momentum and energy applicable
to any law of interaction. By momentum conservation, the velocity
of the center of mass $\boldsymbol{V_{g}}$ is a constant

\[
m_{1}\boldsymbol{v_{1}}+m_{2}\boldsymbol{v_{2}}=m_{1}\boldsymbol{v_{1}'}+m_{2}\boldsymbol{v_{2}'}\equiv\left(m_{1}+m_{2}\right)\boldsymbol{V_{g}}.
\]
Hence we can write, denoting $M_{1}=m_{1}/\left(m_{1}+m_{2}\right)$
and $M_{2}=m_{2}/\left(m_{1}+m_{2}\right)$
\begin{equation}
V_{g}^{2}=M_{1}^{2}v_{1}^{2}+M_{2}^{2}v_{2}^{2}+2M_{1}M_{2}v_{1}v_{2}\cos\theta.\label{eq:VgSqr}
\end{equation}

\begin{figure}
\centering{}\includegraphics[scale=0.9]{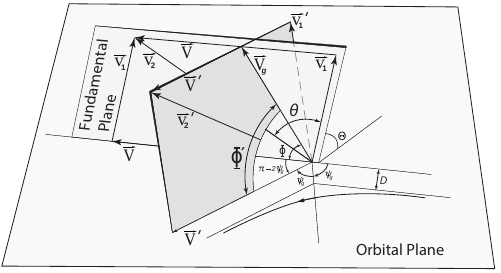}\caption{Velocities for the two particle collision.}
\end{figure}

Let $\boldsymbol{V}\equiv\boldsymbol{v_{2}-v_{1}}$ and $\boldsymbol{V}'\equiv\boldsymbol{v_{2}'-v_{1}'}$
which using the above relations we can write

\begin{equation}
\begin{array}{ccc}
\boldsymbol{v_{1}}=\boldsymbol{V_{g}}-M_{2}\boldsymbol{V}, &  & \boldsymbol{v_{1}'}=\boldsymbol{V_{g}}-M_{2}\boldsymbol{V';}\\
\boldsymbol{v_{2}}=\boldsymbol{V_{g}}+M_{1}\boldsymbol{V}, &  & \boldsymbol{v_{2}'}=\boldsymbol{V_{g}}+M_{2}\boldsymbol{V'.}
\end{array}\label{eq:labVelocities}
\end{equation}
Since the scattering is elastic, the total kinetic energy is conserved

\[
\frac{1}{2}m_{1}v_{1}^{2}+\frac{1}{2}m_{2}v_{2}^{2}=\frac{1}{2}m_{1}v_{1}^{'}\mbox{}^{2}+\frac{1}{2}m_{2}v'_{2}\mbox{}^{2};
\]
it is easy to see that $V=V'$, namely the relative velocity only
changes in direction and \textit{not} magnitude. The dynamical effect
of encounter is therefore known when the change in the direction of
$V$ is determined; hence the importance of the scattering angle.

\subsection{Calculation of $\Delta E$}

By definition $\Delta E=\frac{1}{2}m_{2}v_{2}^{'}\mbox{}^{2}-\frac{1}{2}m_{2}v_{2}^{2}$
and is positive when the test particle gains energy in collision and
is negative when it loses energy. Squaring the quantities in Eq. \ref{eq:labVelocities},
we have

\begin{eqnarray}
v_{2}^{2} & = & V_{g}^{2}+2M_{1}V_{g}V\cos\Phi+M_{1}^{2}V^{2},\label{eq:v2sqrt}\\
v_{1}^{2} & = & V_{g}^{2}-2M_{2}V_{g}V\cos\Phi+M_{2}^{2}V^{2},\label{eq:v1sqr}
\end{eqnarray}
where $\Phi$ is the angle between $\boldsymbol{V_{g}}$ and $\boldsymbol{V}$.
Similarly after the encounter we have

\[
v'_{2}\mbox{}^{2}=V_{g}^{2}+2M_{1}V_{g}V'\cos\Phi'+M_{1}^{2}V'\mbox{}^{2}
\]
where $\Phi'$ is the angle between $\boldsymbol{V_{g}}$ and $\boldsymbol{V'}$.
Since $V=V'$

\begin{eqnarray}
\Delta E & = & \frac{1}{2}m_{2}\left(v'_{2}\mbox{}^{2}-v_{2}^{2}\right)\nonumber \\
 & = & \frac{m_{1}m_{2}}{m_{1}+m_{2}}V_{g}V\left(\cos\Phi'-\cos\Phi\right).
\end{eqnarray}
Solving \ref{eq:v1sqr} and using \ref{eq:VgSqr} and \ref{eq:labVelocities}
for $\cos\Phi$ and using geometry to infer $\cos\Phi'$ we obtain

\begin{eqnarray*}
\cos\Phi & = & \frac{m_{2}v_{2}^{2}-m_{1}v_{1}^{2}+\left(m_{1}-m_{2}\right)v_{1}v_{2}\cos\theta}{\left(m_{1}+m_{2}\right)V_{g}V}\\
\cos\Phi' & = & \cos\Phi\cos\left(\pi-2\psi_{g}\right)+\sin\Phi\sin\left(\pi-2\psi_{g}\right)\cos\Theta.
\end{eqnarray*}
Comment: Let $\mbox{sgn}\alpha$ denote the sign of the interaction;
i.e., $\mbox{sgn}\alpha=+1$ for repulsive and $\mbox{sgn}\alpha=-1$
for attractive interactions. Note that $\sin\left(\pi-2\psi_{g}\right)=\mbox{sgn}\alpha\sin\left(\left|\pi-2\psi_{g}\right|\right)$
.

Using these

\[
\Delta E=-\frac{2m_{1}m_{2}}{m_{1}+m_{2}}V_{g}V\cos^{2}\psi_{g}\left(\cos\Phi-\sin\Phi\cos\Theta\tan\psi_{g}\right).
\]
This can be written in its final form

\begin{equation}
\Delta E=2a\sin\psi_{g}\cos\psi_{g}\cos\Theta-b\cos^{2}\psi_{g},\label{eq:DeltaE}
\end{equation}
where

\begin{eqnarray*}
a & = & \mu v_{1}v_{2}\sin\theta\\
b & = & K_{12}\left[\frac{1}{2}m_{2}v_{2}^{2}-\frac{1}{2}m_{1}v_{1}^{2}+\frac{1}{2}\left(m_{1}-m_{2}\right)v_{1}v_{2}\cos\theta\right],
\end{eqnarray*}
and

\begin{eqnarray*}
\mu & \equiv & \left(m_{1}m_{2}\right)/\left(m_{1}+m_{2}\right),\\
K_{12} & \equiv & 4\left(m_{1}m_{2}\right)/\left(m_{1}+m_{2}\right)^{2}.
\end{eqnarray*}
Below we will use this form of the equations. 

Alternatively, one can expand $\sin\theta$ and $\cos\theta$ in terms
of relative velocity $V$, and write $a$ and $b$  in the form

\begin{eqnarray*}
a & = & \frac{\mu}{2}\left[-V^{4}+2V^{2}\left(v_{1}^{2}+v_{2}^{2}\right)-\left(v_{2}^{2}-v_{1}^{2}\right)^{2}\right]^{1/2}\\
b & = & \mu\left(v_{2}^{2}-v_{1}^{2}+\frac{m_{2}-m_{2}}{m_{2}+m_{1}}V^{2}\right).
\end{eqnarray*}
Similarly equations for $\Delta\tilde{E}$,$\vartheta$,$\phi$, $\tilde{\vartheta}$
,$\tilde{\phi}$ can be obtained \cite{Gryzinski1965_I}.

\section{Geometry and Dynamics Entirely in The Laboratory System}

\subsection{$\boldsymbol{V}$, $D$ and $\Theta$ in laboratory system}

We distinguish between geometrical coordinates and the dynamical coordinates.
By \textit{geometrical coordinates} we have in mind the configuration
of the system when the two particles do not interact and with \textit{dynamical
coordinates} we have in mind the coordinates in the presence of the
force between the particles. In order to express the impact parameter
$D$ and relative velocity $\boldsymbol{V}$ entirely in the laboratory
frame, we work with the geometrical coordinates. 
\begin{figure}
\centering{}\includegraphics[scale=0.28]{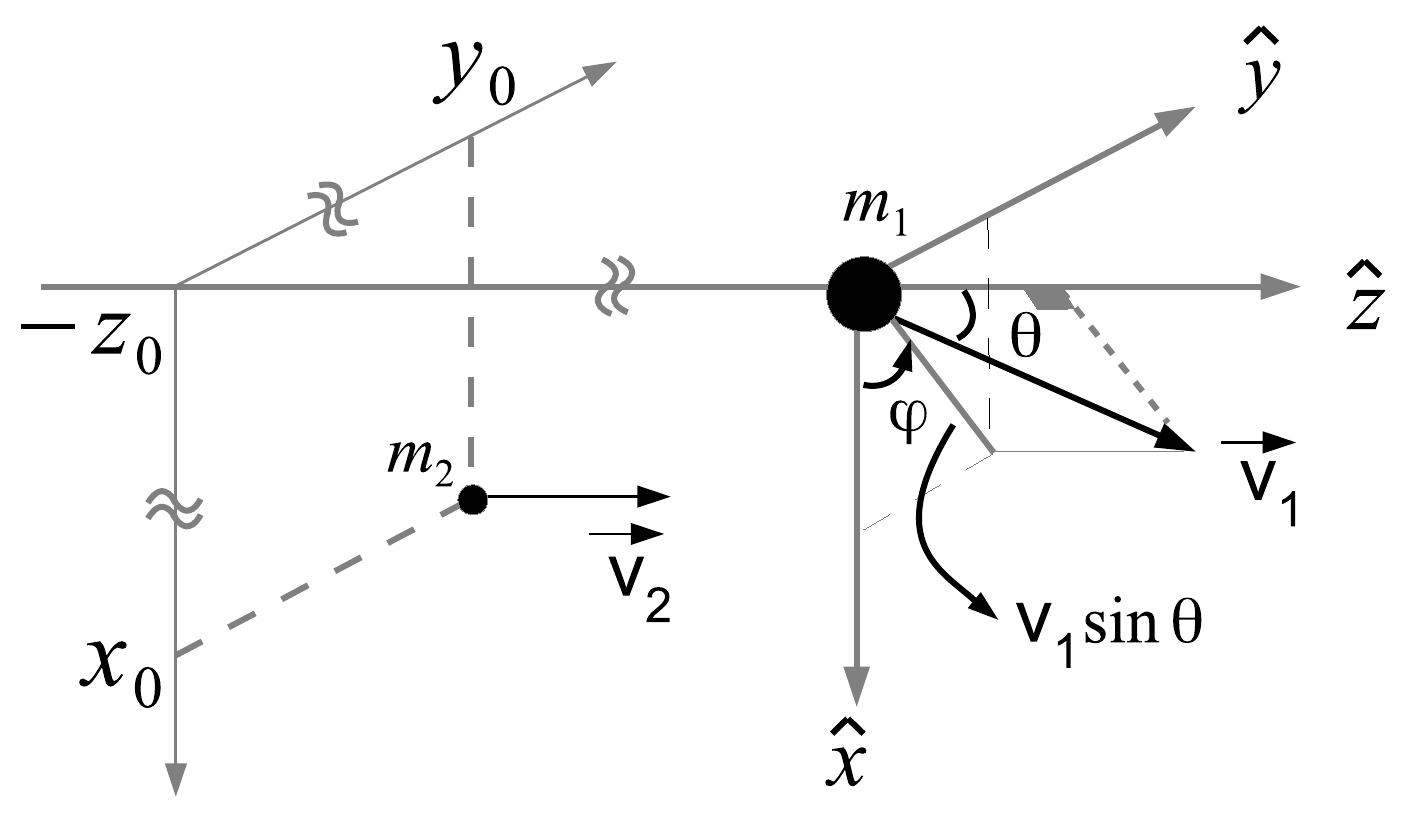}\caption{\label{fig:The-configuration}The initial configuration}
\end{figure}

For any given encounter, let us set up a coordinate system with the
$\boldsymbol{\hat{z}}$ pointing in the direction of $\boldsymbol{v_{2}}$
and put the field particle at its origin. The relative position and
velocity are readily expressed in the lab frame respectively by $\boldsymbol{r}=\boldsymbol{r}_{2}-\boldsymbol{r}_{1}=x_{0}\hat{\boldsymbol{x}}+y_{0}\hat{\boldsymbol{y}}-z_{0}\hat{\boldsymbol{z}}$
and $\boldsymbol{V}=\left(v_{1}\sin\theta\cos\varphi\right)\hat{\boldsymbol{x}}+\left(v_{1}\sin\theta\sin\varphi\right)\hat{\boldsymbol{y}}+\left(v_{1}\cos\theta-v_{2}\right)\hat{\boldsymbol{z}}$
and $V=\left\{ v_{1}^{2}+v_{2}^{2}-2v_{1}v_{2}\cos\theta\right\} ^{1/2}.$
The configuration is shown in  Figure \ref{fig:The-configuration}.

The impact parameter is the distance of closest approach and is obtained
by considering the parametric equations of the lines that each particle
traces and finding the minimum distance between those lines. Let the
line traced by the first particle be denoted by $l_{1}$ and the line
traced by the second particle $l_{2}$. It is necessary to provide
some information regarding the initial configuration by providing
the coordinates of the second particle (first particle being at the
origin in $t=0$). Let the second particle have coordinates $\left(x_{0},y_{0},-z_{0}\right)$,
where we take $z_{0}>0$. Then any point on $l_{1}$ denoted by $P_{1}$
and any point on $l_{2}$ denoted by $P_{2}$ at time $t$ is 

\begin{eqnarray*}
P_{1}\left(t\right) & = & v_{1}t\left(\sin\theta\cos\varphi,\sin\theta\sin\varphi,\cos\theta\right)\\
P_{2}\left(t\right) & = & \left(x_{0},y_{o},-z_{0}+v_{2}t\right).
\end{eqnarray*}
The impact parameter corresponds to the distance between the points
at a time denoted by $t_{*}$ when $\left|P_{2}\left(t\right)-P_{1}\left(t\right)\right|$
is minimized. That time $t_{*}$ is found by $\frac{d}{dt}\left[P_{2}\left(t\right)-P_{1}\left(t\right)\right]^{2}=0$,
where $P_{2}\left(t\right)-P_{1}\left(t\right)=\left(x_{0}-v_{1}t\sin\theta\cos\varphi,y_{0}-v_{1}t\sin\theta\sin\varphi,-z_{0}+v_{2}t-v_{1}t\cos\theta\right)$

\begin{widetext}

\begin{eqnarray*}
\left[P_{2}\left(t\right)-P_{1}\left(t\right)\right]^{2} & = & \left\{ t^{2}\left(v_{1}^{2}+v_{2}^{2}-2v_{1}v_{2}\cos\theta\right)+x_{0}^{2}+y_{0}^{2}+z_{0}^{2}\right.\\
 &  & \left.-2t\left[v_{1}y_{0}\sin\theta\cos\varphi+v_{1}y_{0}\sin\theta\sin\varphi+z_{0}\left(v_{2}-v_{1}\cos\theta\right)\right]\right\} \\
\frac{d}{dt}\left[P_{2}\left(t_{*}\right)-P_{1}\left(t_{*}\right)\right]^{2}=0 & \implies & t_{*}=\frac{v_{1}\sin\theta\left(x_{0}\cos\varphi+y_{0}\sin\varphi\right)+z_{0}\left(v_{2}-v_{1}\cos\theta\right)}{v_{1}^{2}+v_{2}^{2}-2v_{1}v_{2}\cos\theta}.
\end{eqnarray*}

\end{widetext}

Therefore,

\begin{widetext}

\begin{eqnarray}
D & \equiv & \left|P_{2}\left(t_{*}\right)-P_{1}\left(t_{*}\right)\right|\label{eq:D_lab_Exact}\\
 & = & \left\{ x_{0}^{2}+y_{0}^{2}-\frac{v_{1}\sin\theta}{V^{2}}\left[\frac{v_{1}}{2}\sin\theta\left(x_{0}^{2}+y_{0}^{2}-2z_{0}^{2}+\left(x_{0}^{2}-y_{0}^{2}\right)\cos2\varphi+2x_{0}y_{0}\sin2\varphi\right)\right.\right.\nonumber \\
 &  & \left.\left.+2v_{2}z_{0}\left(1-\frac{v_{1}}{v_{2}}\cos\theta\right)\left(x_{0}\cos\varphi+y_{0}\sin\varphi\right)\right]\right\} ^{1/2}\nonumber 
\end{eqnarray}

\end{widetext}

$\Theta$ is the angle between the fundamental and orbital planes
and is defined by $\cos\Theta=\boldsymbol{\hat{n}_{orb}}\cdot\boldsymbol{\hat{n}_{F}}$,
where $\boldsymbol{\hat{n}_{orb}}=\frac{\boldsymbol{r}\times\boldsymbol{V}}{\left|\boldsymbol{r\times}\boldsymbol{V}\right|}$
and $\boldsymbol{\hat{n}_{F}}\equiv\frac{\boldsymbol{v_{1}\times}\boldsymbol{v_{2}}}{v_{1}v_{2}\sin\theta}$
are the unit vectors perpendicular to the orbital and the fundamental
planes respectively
\begin{eqnarray*}
\boldsymbol{\hat{n}_{F}} & = & \left(\sin\varphi,-\cos\varphi,0\right),\\
\boldsymbol{\hat{n}_{orb}} & = & \frac{1}{\left|\boldsymbol{r\times}\boldsymbol{V}\right|}\left(n_{orb}^{x},n_{orb}^{y},n_{orb}^{z}\right),\\
n_{orb}^{x} & \equiv & y_{0}\left(v_{2}-v_{1}\cos\theta\right)-v_{1}z_{0}\sin\theta\sin\varphi,\\
n_{orb}^{y} & \equiv & -v_{2}x_{0}+v_{1}\left(x_{0}\cos\theta+z_{0}\sin\theta\cos\varphi\right),\\
n_{orb}^{z} & \equiv & v_{1}\sin\theta\left(y_{0}\cos\varphi-x_{0}\sin\varphi\right),
\end{eqnarray*}
which gives

\begin{eqnarray*}
\cos\Theta & = & \frac{\left(x_{0}\cos\varphi+y_{0}\sin\varphi\right)\left(v_{2}-v_{1}\cos\theta\right)-v_{1}z_{0}\sin\theta}{B}\\
B & \equiv & \left\{ \left[v_{2}x_{0}-v_{1}\left(x_{0}\cos\theta+z_{0}\sin\theta\cos\varphi\right)\right]^{2}\right.\\
 & + & v_{1}^{2}\sin^{2}\theta\left(-x_{0}\sin\varphi+y_{0}\cos\varphi\right)^{2}\\
 & + & \left.\left[y_{0}\left(v_{2}-v_{1}\cos\theta\right)-v_{1}z_{0}\sin\theta\sin\varphi\right]^{2}\right\} ^{1/2}.
\end{eqnarray*}

The orbital and fundamental planes coincide for $y_{0}=0$, $\theta=\frac{\pi}{2}$
and $\varphi=0,\pi$. We see that in these special cases taking $y_{0}=0$
and $\theta=\frac{\pi}{2}$

\begin{eqnarray}
\cos\Theta & = & \frac{x_{0}v_{2}-v_{1}z_{0}}{\sqrt{\left(v_{2}x_{0}-v_{1}z_{0}\right)^{2}}}=1\qquad\varphi=0\label{eq:CosTheta_approach}\\
\cos\Theta & = & \frac{-x_{0}v_{2}-v_{1}z_{0}}{\sqrt{\left(v_{2}x_{0}+v_{1}z_{0}\right)^{2}}}=-1\qquad\varphi=\pi.\label{eq:cosTheta_recede}
\end{eqnarray}

\subsection{\label{sub:Effective-Collision-Times}Effective Collision Times}

The discussions so far refer to ideal scattering processes where the
particles start infinitely apart and go to infinity after the collision
takes place. The scattering angle $\psi_{g}$ is the angle between
the two asymptotes. In real collisions, as the ones being considered
here, the test particle scatters from many field particles that are
far yet at finite distances from one another. Therefore, $\psi_{g}$
over-estimates the actual scattering angle per collision. Since the
entire effect of interaction is uniquely determined by $\psi_{g}$,
we let define the collision time to be the time after which this angle
attains a value close to the value it would have attained in infinite
time. In the case of central forces under consideration Gryzinski
\cite[Section IV]{Gryzinski1965_I} divides the collisions to two
types by defining a parameter $r_{0}$ to be the distance at which
the potential energy of the two particles is equal to the relative
kinetic energy. The collisions with an impact parameter $D<r_{0}$
are called the ``close collisions'' with $D>r_{0}$ are called ``distant
collisions''. The collision time for central forces with potential
$U\left(r\right)=\frac{\alpha}{r^{k}}$ are shown to be \cite{Gryzinski1965_I}

\[
t_{coll}\approx2\left[\left(r_{0}+D\right)/V\right]\left(2^{1/k}-1\right),
\]
 which for the Coulomb interaction $k=1$ becomes $t_{coll}\approx2\left[\left(r_{0}+D\right)/V\right]$. 

In addition, for a series of two-body scatterings to be a sensible
approximation of the many-body phenomena the time of collision needs
to be much shorter than the time it takes for the test or field particle
to have an appreciable change in their velocities due to interaction
with other particles or external fields.

\subsection{Small Angle Scattering}

So far the formulation has been exact. For potentials of type $\alpha/r^{k}$
a small angle scattering corresponds to $D\gg\frac{\alpha}{\mu V^{2}}$.
We show this for $k=1,2$ and note that for $k>2$ the field strength
decreases stronger with the distance and the same condition $D\gg\frac{\alpha}{\mu V^{2}}$
ought to suffice. This can be seen from Eqs. \ref{eq:cosPsi_k=00003D1}
- \ref{eq:Psi_k=00003D2} which in this limit read

\begin{eqnarray*}
\sin\psi_{g} & \underset{=}{\left(k=1\right)} & \frac{1}{\left\{ 1+\left(\frac{\alpha}{\mu DV^{2}}\right)^{2}\right\} ^{1/2}}\approx\left\{ 1-\frac{1}{2}\left(\frac{\alpha}{\mu DV^{2}}\right)^{2}\right\} ,\\
\cos\psi_{g} & \underset{=}{\left(k=1\right)} & \left(\frac{\alpha}{\mu DV^{2}}\right)\left\{ 1+\left(\frac{\alpha}{\mu DV^{2}}\right)^{2}\right\} ^{-1/2}\approx\frac{\alpha}{\mu DV^{2}},\\
\psi_{g} & \underset{=}{\left(k=2\right)} & \frac{\pi}{2}\frac{1}{\left\{ 1+\left(\frac{2\alpha}{\mu V^{2}D^{2}}\right)\right\} ^{1/2}}\approx\frac{\pi}{2}\left[1-\left(\frac{\alpha}{\mu D^{2}V^{2}}\right)\right].
\end{eqnarray*}

Since $\Delta E$ depends on it, we approximate $\sin\psi_{g}\cos\psi_{g}\approx\frac{\alpha}{\mu DV^{2}}\left[1-\left(\frac{\alpha}{\mu DV^{2}}\right)^{2}\right].$
From now on we restrict ourselves to the important case of $k=1$.

Comment: There are two small parameters under consideration. 1. $\frac{\alpha}{\mu V^{2}}\ll D$
which allows us to approximate the dynamical quantities and 2. $\frac{v_{1}}{v_{2}}$
that we use for approximating the geometric quantities. We shall see
that terms of order $\left(\frac{v_{1}}{v_{2}}\right)^{2}$ are necessary
to keep to obtain the asymmetry we seek in the ensemble average $\langle\Delta E\rangle_{\boldsymbol{v_{1}}}$.

Below we keep to second order in $\frac{\alpha}{\mu DV^{2}}$.

\vspace{0.4cm}

\section{Statistics: Ensemble Average}

What needs to be done for our purposes is to calculate the \textit{ensemble}
average $\langle\Delta E\rangle_{\boldsymbol{v_{1}}}$, where by the
subscript we have in mind average with respect to random variable
$\boldsymbol{v_{1}}$. To bring out the effect first let us fix the
speed $v_{1}$ and let $f\left(\theta,\varphi\right)d\theta d\varphi$
denote the probability that the velocity vector of $m_{1}$ has direction
determined by angles $\theta$ and $\varphi$. For isotropically moving
field particles $f\left(\theta,\varphi\right)=\frac{1}{2}\sin\theta\left(\frac{1}{2\pi}\right)$.
It is found under rather general conditions that the distribution
function of the speed of the field stars is given by \cite[Eq: 2.353]{chandra4}

\[
N\left(v_{1}\right)dv_{1}=\frac{4j^{3}}{\sqrt{\pi}}Ne^{-j^{2}v_{1}^{2}}v_{1}^{2}dv_{1}.
\]
The measure for the ensemble average then becomes

\begin{equation}
f\left(\theta,\varphi\right)N\left(v_{1}\right)dv_{1}d\theta d\varphi=\frac{Nj^{3}}{\pi^{3/2}}\sin\theta e^{-j^{2}v_{1}^{2}}v_{1}^{2}dv_{1}d\theta d\varphi\label{eq:measure}
\end{equation}

\subsection{$v_{1}\ll v_{2}$ and small angle scattering and $m_{2}\sim m_{1}$}

When the masses are comparable one expects that $m_{2}$ on average
would impart energy to $m_{1}$. We include this case as a side calculation
because of its simplicity and relevance for phenomena beyond the scope
of this work. From above we have

\[
\Delta E=2a\sin\psi_{g}\cos\psi_{g}\cos\Theta-b\cos^{2}\psi_{g},
\]
where $\mu=\frac{m_{1}m_{2}}{m_{1}+m_{2}}$ and $K_{12}\equiv\frac{4\mu}{m_{1}+m_{2}}$

\begin{eqnarray*}
a & = & \mu v_{1}v_{2}\sin\theta\\
b & = & K_{12}\left[\frac{1}{2}m_{2}v_{2}^{2}-\frac{1}{2}m_{1}v_{1}^{2}+\frac{1}{2}\left(m_{1}-m_{2}\right)v_{1}v_{2}\cos\theta\right]
\end{eqnarray*}

\begin{eqnarray*}
\frac{1}{4\pi}\int d\theta d\varphi\sin\theta\Delta E & = & \frac{1}{4\pi}\int d\theta d\varphi\sin\theta\left\{ 2a\sin\psi_{g}\cos\psi_{g}\cos\Theta\right\} \\
 & - & \frac{1}{4\pi}\int d\theta d\varphi\sin\theta\left\{ b\cos^{2}\psi_{g}\right\} 
\end{eqnarray*}

To zeroth order in $\frac{v_{1}}{v_{2}}$ we find $\sin\psi_{g}\approx\frac{\mu v_{2}^{2}D_{0}}{\sqrt{1+\frac{\left(\mu v_{2}^{2}D_{0}\right)^{2}}{\alpha}}}$
and $\cos\psi_{g}\approx\left\{ 1+\left(\frac{\mu v_{2}^{2}D_{0}}{\alpha}\right)^{2}\right\} ^{-1/2}$.
Consequently to zeroth order we find $2a\sin\psi_{g}\cos\psi_{g}\cos\Theta\approx0$
and $b\cos^{2}\psi_{g}\approx\frac{2m_{1}\left(m_{2}v_{2}\alpha\right)^{2}}{\left(m_{1}m_{2}v_{2}^{2}D_{0}\right)^{2}+\left(m_{1}+m_{2}\right)^{2}\alpha^{2}}$,
which readily gives us a first order effect

\[
\langle\Delta E\rangle_{\theta,\varphi}=-\frac{2m_{1}\left(m_{2}v_{2}\alpha\right)^{2}}{\left(m_{1}m_{2}v_{2}^{2}D_{0}\right)^{2}+\left(m_{1}+m_{2}\right)^{2}\alpha^{2}}+\mathcal{O}\left(\frac{v_{1}}{v_{2}}\right).
\]
Note that we did \textit{not} make approximations using $m_{2}\ll m_{1}$.
We see that regardless of the sign of the interaction, the condition
$v_{1}\ll v_{2}$ implies that the fast particle on average must lose
energy to the slower one.

\subsection{Approximation of dynamical quantities via: $m_{2}\ll m_{1}$ , $v_{2}\gg v_{1}$
but $m_{1}v_{1}^{2}\gg m_{2}v_{2}^{2}$ }

The relation $D\gg\frac{\alpha}{\mu V^{2}}$ can be satisfied in various
ways. We are interested in glazing collisions of $\boldsymbol{v_{2}}$
from $\boldsymbol{v_{1}}$ such that $v_{1}\ll v_{2}$ and $m_{2}\ll m_{1}$
but $m_{1}v_{1}^{2}\gg m_{2}v_{2}^{2}$. These conditions together
imply $m_{1}v_{1}\gg m_{2}v_{2}$. The relative momenta are important
for our purposes of calculating the \textit{ensemble} average $\langle\Delta E\rangle_{\boldsymbol{v_{1}}}$,
where by the subscript we have in mind average with respect to random
variable $\boldsymbol{v_{1}}$. From above we have

\[
\Delta E=2a\sin\psi_{g}\cos\psi_{g}\cos\Theta-b\cos^{2}\psi_{g},
\]
which in the limit is specified by 

\begin{eqnarray}
a & = & \mu v_{1}v_{2}\sin\theta\approx m_{2}v_{1}v_{2}\sin\theta\label{eq:a_firstOrder}\\
b & = & K_{12}\left[\frac{1}{2}m_{2}v_{2}^{2}-\frac{1}{2}m_{1}v_{1}^{2}+\frac{1}{2}\left(m_{1}-m_{2}\right)v_{1}v_{2}\cos\theta\right]\nonumber \\
 & \approx & 2m_{2}v_{2}^{2}\left(\frac{v_{1}}{v_{2}}\right)\left(\cos\theta-\frac{v_{1}}{v_{2}}\right).\label{eq:b_firstOrder}
\end{eqnarray}
where $\mu=\frac{m_{1}m_{2}}{m_{1}+m_{2}}$ and $K_{12}\equiv\frac{4\mu}{m_{1}+m_{2}}$.
In order to calculate $\Delta E$ we need to approximate the geometric
quantities.

\subsection{Approximation of geometrical quantities to first order in $\left(\frac{v_{1}}{v_{2}}\right)$ }

The condition $v_{1}\ll v_{2}$ is enough to allow approximations
of the geometrical coordinates. The inertia of the particles and the
strength of the interaction are irrelevant in calculation of the geometric
coordinates. 

Here to make appropriate approximations we assume $x_{0},y_{o},z_{0}$
are of the same order of magnitude (see Figure \ref{fig:The-configuration}).
We then calculate the ensemble average ignoring terms of $\mathcal{O}\left(\frac{v_{1}}{v_{2}}\right)^{2}$
and higher

\begin{eqnarray*}
V^{2} & \approx & v_{2}^{2}\left(1-\frac{2v_{1}}{v_{2}}\cos\theta\right)\\
v_{z} & = & v_{2}\left(1-\frac{v_{1}}{v_{2}}\cos\theta\right)\\
D & \approx & \left\{ x_{0}^{2}+y_{0}^{2}-\frac{2v_{1}}{v_{2}}z_{0}\sin\theta\left(x_{0}\cos\varphi+y_{0}\sin\varphi\right)\right\} ^{1/2}\\
 & \approx & D_{0}\left\{ 1-\frac{v_{1}}{v_{2}}\frac{z_{0}}{D_{0}}\sin\theta\left(\frac{x_{0}}{D_{0}}\cos\varphi+\frac{y_{0}}{D_{0}}\sin\varphi\right)\right\} ,
\end{eqnarray*}
where $D_{0}\equiv\sqrt{x_{0}^{2}+y_{0}^{2}}$. Lastly,

\begin{eqnarray*}
\cos\Theta & \approx & -\frac{x_{0}\cos\varphi+y_{0}\sin\varphi}{D_{0}}\\
 & + & \frac{z_{0}\sin\theta}{D_{0}}\left(1-\frac{x_{0}y_{0}\sin2\varphi}{D_{0}^{2}}\right)\left(\frac{v_{1}}{v_{2}}\right).
\end{eqnarray*}

Similarly to first order 

\begin{widetext}

\begin{eqnarray*}
\sin\psi_{g} & \underset{=}{\left(k=1\right)} & \frac{1}{\left\{ 1+\left(\frac{\alpha}{\mu DV^{2}}\right)^{2}\right\} ^{1/2}}\approx\left\{ 1-\frac{1}{2}\left(\frac{\alpha}{\mu D_{0}v_{2}^{2}}\right)^{2}\left(1+\frac{v_{1}}{v_{2}}\left(\cos\theta+2\sin\theta\sin\varphi\right)\right)\right\} \\
\cos\psi_{g} & \underset{\approx}{\left(k=1\right)} & \frac{\alpha}{\mu D_{0}v_{2}^{2}}\left\{ 1+\left(\frac{v_{1}}{v_{2}}\right)\left[2\cos\theta+\frac{z_{0}}{D_{0}}\sin\theta\left(\frac{x_{0}}{D_{0}}\cos\varphi+\frac{y_{0}}{D_{0}}\sin\varphi\right)\right]\right\} \\
\sin\psi_{g}\cos\psi_{g} & \underset{\approx}{\left(k=1\right)} & \frac{\alpha}{\mu DV^{2}}\approx\frac{\alpha}{\mu D_{0}v_{2}^{2}}\left\{ 1+\left(\frac{v_{1}}{v_{2}}\right)\left[2\cos\theta+\frac{z_{0}}{D_{0}}\sin\theta\left(\frac{x_{0}}{D_{0}}\cos\varphi+\frac{y_{0}}{D_{0}}\sin\varphi\right)\right]\right\} .
\end{eqnarray*}

\end{widetext}

We can examine $\Delta E$ for the ``transverse'' collisions to
first order in $\frac{v_{!}}{v_{2}}$. It suffices to consider $y_{0}=0$,
whereby $D_{0}=x_{0}$ and $\theta=\frac{\pi}{2}$. In the approaching
case $\varphi=0$ and in the receding case $\varphi=\pi$ (See Figure
\ref{fig:The-configuration}). $\cos\Theta$ was obtained exactly
for these special case in Eqs. \ref{eq:CosTheta_approach} and \ref{eq:cosTheta_recede}.
Further for $\theta=\frac{\pi}{2}$ we have

\begin{widetext}

\begin{eqnarray*}
\sin\psi_{g}\cos\psi_{g} & \approx & \frac{\alpha}{\mu D_{0}v_{2}^{2}}\left\{ 1+\left(\frac{v_{1}}{v_{2}}\right)\left[\frac{z_{0}}{D_{0}}\left(\frac{x_{0}}{D_{0}}\cos\varphi+\frac{y_{0}}{D_{0}}\sin\varphi\right)\right]\right\} \\
\cos^{2}\psi_{g} & \approx & \left(\frac{\alpha}{\mu D_{0}v_{2}^{2}}\right)^{2}\left\{ 1+\left(\frac{v_{1}}{v_{2}}\right)\left[2\cos\theta+\frac{z_{0}}{D_{0}}\sin\theta\left(\frac{x_{0}}{D_{0}}\cos\varphi+\frac{y_{0}}{D_{0}}\sin\varphi\right)\right]\right\} ^{2}\\
 & \approx & \left(\frac{\alpha}{\mu D_{0}v_{2}^{2}}\right)^{2}\left\{ 1+2\left(\frac{v_{1}}{v_{2}}\right)\left[2\cos\theta+\frac{z_{0}}{D_{0}}\sin\theta\left(\frac{x_{0}}{D_{0}}\cos\varphi+\frac{y_{0}}{D_{0}}\sin\varphi\right)\right]\right\} 
\end{eqnarray*}

\end{widetext}

These combined with Eqs. \ref{eq:a_firstOrder} and \ref{eq:b_firstOrder}
give for

\textbf{Approach case:} $\theta=\frac{\pi}{2}$ and $\varphi=0$,
$\cos\Theta=1$: $\Delta E=2a\sin\psi_{g}\cos\psi_{g}-b\cos^{2}\psi_{g}$
, keeping to first order

\begin{eqnarray*}
\Delta E_{approach} & = & \frac{2\alpha}{D_{0}}\frac{v_{1}}{v_{2}}\left\{ 1+\left(\frac{v_{1}}{v_{2}}\right)\left[\frac{z_{0}}{D_{0}}\left(\frac{x_{0}}{D_{0}}\right)\right]\right\} \\
 & + & 2m_{2}\left(\frac{v_{1}}{v_{2}}\right)^{2}\left(\frac{\alpha}{\mu D_{0}v_{2}}\right)^{2}\\
 & \times & \left\{ 1+2\left(\frac{v_{1}}{v_{2}}\right)\left[\frac{z_{0}}{D_{0}}\left(\frac{x_{0}}{D_{0}}\right)\right]\right\} \\
 & = & \frac{2\alpha}{D_{0}}\frac{v_{1}}{v_{2}}+\mathcal{O}\left(\frac{v_{1}}{v_{2}}\right)^{2}\\
 & = & \left\{ \begin{array}{c}
<0\quad\mbox{ attractive}\quad\alpha<0\\
>0\quad\mbox{ repulsive }\quad\alpha>0
\end{array}\right.
\end{eqnarray*}

Moreover, we see that in the approaching case $\cos\psi_{g}^{a}=\frac{\alpha}{\mu D_{0}v_{2}^{2}}\left\{ 1+\left(\frac{v_{1}}{v_{2}}\right)\left[\frac{z_{0}}{D_{0}}\left(\frac{x_{0}}{D_{0}}\right)\right]\right\} $.

\textbf{Recede case:} $\theta=\frac{\pi}{2}$ and $\varphi=\pi$,
$\cos\Theta=-1$: $\Delta E=-\left(2a\sin\psi_{g}\cos\psi_{g}+b\cos^{2}\psi_{g}\right)$ 

\begin{eqnarray*}
\Delta E_{recede} & = & -\frac{2\alpha}{D_{0}}\frac{v_{1}}{v_{2}}\left\{ 1+\left(\frac{v_{1}}{v_{2}}\right)\left[\frac{z_{0}}{D_{0}}\left(\frac{x_{0}}{D_{0}}\right)\right]\right\} \\
 & + & 2m_{2}\left(\frac{v_{1}}{v_{2}}\right)^{2}\left(\frac{\alpha}{\mu D_{0}v_{2}}\right)^{2}\\
 & \times & \left\{ 1+2\left(\frac{v_{1}}{v_{2}}\right)\left[\frac{z_{0}}{D_{0}}\left(\frac{x_{0}}{D_{0}}\right)\right]\right\} \\
 & = & -\frac{2\alpha}{D_{0}}\frac{v_{1}}{v_{2}}+\mathcal{O}\left(\frac{v_{1}}{v_{2}}\right)^{2}\\
 & = & \left\{ \begin{array}{c}
>0\quad\mbox{ attractive}\quad\alpha<0\\
<0\quad\mbox{repulsive }\quad\alpha>0
\end{array}\right.
\end{eqnarray*}

Moreover, we see that in the receding case $\cos\psi_{g}^{r}=\frac{\alpha}{\mu D_{0}v_{2}^{2}}\left\{ 1-\left(\frac{v_{1}}{v_{2}}\right)\left[\frac{z_{0}}{D_{0}}\left(\frac{x_{0}}{D_{0}}\right)\right]\right\} <\cos\psi_{g}^{a}$,
proving our assertion in the paper. 

Since the two extreme cases do \textit{not} show any asymmetry to
first order in $\frac{v_{1}}{v_{2}}$ we expect $\langle\Delta E\rangle_{\boldsymbol{v_{1}}}=0$.
To bring out the effect first let us fix the speed $v_{1}$ and let
$f\left(\theta,\varphi\right)d\theta d\varphi$ denote the probability
that the velocity vector of $m_{1}$ has direction determined by angles
$\theta$ and $\varphi$. For isotropically moving field particles
$f\left(\theta,\varphi\right)=\frac{1}{2}\sin\theta\left(\frac{1}{2\pi}\right)$.
We now prove this by calculating the ensemble average (over $0\le\varphi\le2\pi$
and $0\le\theta\le\pi$) and noting that $b\cos\psi_{g}^{2}=\mathcal{O}\left(\frac{v_{1}}{v_{2}}\right)^{2}$

\begin{widetext}

\begin{eqnarray*}
\frac{1}{4\pi}\int_{0}^{\pi}d\theta\int_{0}^{2\pi}d\varphi\sin\theta\Delta E & = & \frac{1}{4\pi}\int d\theta d\varphi\sin\theta\left\{ \mu v_{1}v_{2}\sin\theta\left(2\sin\psi_{g}\cos\psi_{g}\cos\Theta\right)\right\} \\
 & \approx & \frac{\alpha\mu}{2\pi\mu D_{0}}\frac{v_{1}}{v_{2}}\int d\theta d\varphi\sin^{2}\theta\left\{ 1+\left(\frac{v_{1}}{v_{2}}\right)\left[2\cos\theta+\frac{z_{0}\sin\theta}{D_{0}}\left(\frac{x_{0}\cos\varphi+y_{0}\sin\varphi}{D_{0}}\right)\right]\right\} \\
 & \times & \left\{ -\frac{x_{0}\cos\varphi+y_{0}\sin\varphi}{D_{0}}+\frac{z_{0}\sin\theta}{D_{0}}\left(1-\frac{x_{0}y_{0}\sin2\varphi}{D_{0}^{2}}\right)\left(\frac{v_{1}}{v_{2}}\right)\right\} \\
 & = & \frac{\alpha\mu}{2\pi\mu D_{0}}\frac{v_{1}}{v_{2}}\int d\theta d\varphi\sin^{2}\theta\left(-\frac{x_{0}\cos\varphi+y_{0}\sin\varphi}{D_{0}}\right)+\mathcal{O}\left(\frac{v_{1}}{v_{2}}\right)^{2}\\
 & = & \mathcal{O}\left(\frac{v_{1}}{v_{2}}\right)^{2}
\end{eqnarray*}

\end{widetext}

Some conclusions can be drawn to first order in $\frac{v_{1}}{v_{2}}$:
\begin{enumerate}
\item For $\alpha<0$ the transverse collisions lead to small yet nonzero
energy gain (loss) for the receding (approaching) collisions. For
$\alpha>0$ the transverse collisions lead to small yet nonzero energy
gain (loss) for the approaching (receding) collisions. This proves
the heuristic arguments we gave earlier. 
\item We see that $\cos\psi_{g}^{r}<\cos\psi_{g}^{a}$ as expected when
the field particle approaches the test particle and as a result of
the nonlinearity in the force field breaks the symmetry between the
two cases. 
\item To investigate any asymmetry in $\Delta E$ one needs to go beyond
$\mathcal{O}\left(\frac{v_{1}}{v_{2}}\right)$. 
\end{enumerate}

\subsection{Approximations to second order in $\left(\frac{v_{1}}{v_{2}}\right)$ }

As before $a=m_{2}v_{1}v_{2}\sin\theta$ and $b=2m_{2}v_{1}v_{2}\left[\cos\theta-\left(\frac{v_{1}}{v_{2}}\right)\right]$.
The quantity $2a\sin\psi_{g}\cos\psi_{g}$ is first order in $\left(\frac{v_{1}}{v_{2}}\right)$ 

\begin{widetext}

\[
2a\sin\psi_{g}\cos\psi_{g}\approx\frac{2\alpha\sin\theta}{D_{0}}\frac{v_{1}}{v_{2}}\left\{ 1+\left(\frac{v_{1}}{v_{2}}\right)\left[2\cos\theta+\frac{z_{0}}{D_{0}}\sin\theta\left(\frac{x_{0}}{D_{0}}\cos\varphi+\frac{y_{0}}{D_{0}}\sin\varphi\right)\right]\right\} +\mathcal{O}\left(\frac{v_{1}}{v_{2}}\right)^{3},
\]

\end{widetext}

therefore it is sufficient to approximate $\cos\Theta$ to first order
as well 

\begin{eqnarray*}
\cos\Theta & \approx & -\frac{x_{0}\cos\varphi+y_{0}\sin\varphi}{D_{0}}\\
 & + & \left(\frac{v_{1}}{v_{2}}\right)\frac{z_{0}\sin\theta}{D_{0}}\left(1-\frac{x_{0}y_{0}\sin2\varphi}{D_{0}^{2}}\right).
\end{eqnarray*}

Expanding and ignoring terms of order $\mathcal{O}\left(\frac{v_{1}}{v_{2}}\right)^{3}$
and higher

\begin{widetext}

\begin{eqnarray*}
D & \approx & D_{0}\left\{ 1-\frac{v_{1}}{v_{2}}\frac{z_{0}}{D_{0}}\sin\theta\left(\frac{x_{0}}{D_{0}}\cos\varphi+\frac{y_{0}}{D_{0}}\sin\varphi\right)-\left(\frac{v_{1}}{v_{2}}\right)^{2}\sin\theta\left\{ \frac{z_{0}}{D_{0}}\cos\theta\left(\frac{x_{0}}{D_{0}}\cos\varphi+\frac{y_{0}}{D_{0}}\sin\varphi\right)\right.\right.\\
 & + & \frac{\sin\theta}{4}\left.\left[\left(1-\frac{z_{0}^{2}}{D_{0}^{2}}\right)+\left(1+\frac{z_{0}^{2}}{D_{0}^{2}}\right)\left(\left(\frac{x_{0}^{2}}{D_{0}^{2}}-\frac{y_{0}^{2}}{D_{0}^{2}}\right)\cos2\varphi+\frac{2x_{0}y_{0}}{D_{0}^{2}}\sin2\varphi\right)\right]\right\} \\
V^{2} & = & v_{2}^{2}\left(1-\frac{2v_{1}}{v_{2}}\cos\theta+\left(\frac{v_{1}}{v_{2}}\right)^{2}\right)\\
v_{z} & = & v_{2}\left(1-\frac{v_{1}}{v_{2}}\cos\theta\right).
\end{eqnarray*}

\end{widetext}

We can now calculate $\langle\Delta E\rangle_{\theta,\varphi}=\langle2a\sin\psi_{g}\cos\psi_{g}\cos\Theta-b\cos^{2}\psi_{g}\rangle_{\theta,\varphi}$
term by term where as before the ensemble average is over $0\le\varphi\le2\pi$
and $0\le\theta\le\pi$

\begin{widetext}

\begin{eqnarray*}
\langle a\sin2\psi_{g}\cos\Theta\rangle_{\theta,\varphi} & \approx & \frac{1}{4\pi}\frac{2\alpha}{D_{0}}\left(\frac{v_{1}}{v_{2}}\right)\int d\theta d\varphi\sin^{2}\theta\left\{ 1+\left(\frac{v_{1}}{v_{2}}\right)\left[2\cos\theta+\frac{z_{0}\sin\theta}{D_{0}}\left(\frac{x_{0}\cos\varphi+y_{0}\sin\varphi}{D_{0}}\right)\right]\right\} \\
 & \times & \left\{ -\left[\frac{x_{0}}{D_{0}}\cos\varphi+\frac{y_{0}}{D_{0}}\sin\varphi\right]+\frac{v_{1}}{v_{2}}\frac{z_{0}}{D_{0}}\sin\theta\left(1-\frac{x_{0}y_{0}\sin2\varphi}{D_{0}^{2}}\right)\right\} \\
 & = & \frac{1}{4\pi}\frac{2\alpha}{D_{0}}\left(\frac{v_{1}}{v_{2}}\right)\int_{0}^{\pi}d\theta d\varphi\sin^{2}\theta\left\{ \frac{v_{1}}{v_{2}}\frac{z_{0}}{D_{0}}\sin\theta-\frac{v_{1}}{v_{2}}\frac{z_{0}}{D_{0}}\sin\theta\left(\frac{x_{0}}{D_{0}}\cos\varphi+\frac{y_{0}}{D_{0}}\sin\varphi\right)^{2}\right\} \\
 & = & \frac{1}{2\pi}\frac{\alpha}{D_{0}}\left(\frac{v_{1}}{v_{2}}\right)\left\{ \frac{4}{3}\frac{v_{1}}{v_{2}}\frac{z_{0}}{D_{0}}2\pi-\frac{4}{3}\frac{v_{1}}{v_{2}}\frac{z_{0}}{D_{0}}\pi\right\} =\frac{2}{3}\frac{\alpha}{D_{0}}\frac{z_{0}}{D_{0}}\left(\frac{v_{1}}{v_{2}}\right)^{2}.\\
\end{eqnarray*}

\end{widetext}

Furthermore,

\begin{widetext}

\begin{eqnarray*}
-\langle b\cos^{2}\psi_{g}\rangle_{\theta,\varphi} & = & -\frac{1}{4\pi}\left(2m_{2}\frac{v_{1}}{v_{2}}\right)\left(\frac{\alpha}{\mu D_{0}v_{2}}\right)^{2}\int d\varphi d\theta\left\{ \sin\theta\left(\cos\theta-\frac{v_{1}}{v_{2}}\right)\right.\\
 & \times & \left.\left[1+2\frac{v_{1}}{v_{2}}\left(2\cos\theta+\frac{z_{0}}{D_{0}}\sin\theta\left(\frac{x_{0}}{D_{0}}\cos\varphi+\frac{y_{0}}{D_{0}}\sin\varphi\right)\right)\right]\right\} \\
 & = & -\frac{2}{3m_{2}v_{2}^{2}}\left(\frac{\alpha}{D_{0}}\right)^{2}\left(\frac{v_{1}}{v_{2}}\right)^{2}.
\end{eqnarray*}

\end{widetext}

Therefore we conclude that

\begin{widetext}

\begin{equation}
\langle\Delta E\rangle_{\theta,\varphi}=\frac{2}{3}\frac{\alpha}{D_{0}}\left(\frac{v_{1}}{v_{2}}\right)^{2}\left[\frac{z_{0}}{D_{0}}-\frac{1}{m_{2}v_{2}^{2}}\left(\frac{\alpha}{D_{0}}\right)\right]\approx\frac{2}{3}\frac{\alpha}{D_{0}}\frac{z_{0}}{D_{0}}\left(\frac{v_{1}}{v_{2}}\right)^{2};\label{eq:EnsembleAvg_theta_phi}
\end{equation}

\end{widetext}

because small angle scattering requires that $\frac{1}{m_{2}v_{2}^{2}}\left(\frac{\alpha}{D_{0}}\right)\ll1$
yet $z_{0}$ is comparable to $D_{0}$.

It is found under rather general conditions that the distribution
function of the speed of the field stars is given by \cite[Eq: 2.353]{chandra4}

\[
N\left(v_{1}\right)dv_{1}=\frac{4j^{3}}{\sqrt{\pi}}Ne^{-j^{2}v_{1}^{2}}v_{1}^{2}dv_{1}
\]
where $N$ is the number of field particles per unit volume. Using
this we conclude

\begin{eqnarray}
\langle\Delta E\rangle_{\boldsymbol{v_{1}}} & = & \int N\left(v_{1}\right)dv_{1}\langle\Delta E\rangle_{\theta,\varphi}\nonumber \\
 & = & \frac{2}{3}\frac{\alpha}{D_{0}}\frac{z_{0}}{D_{0}}\frac{4j^{3}}{\sqrt{\pi}}\frac{N}{v_{2}^{2}}\int e^{-j^{2}v_{1}^{2}}v_{1}^{4}dv_{1}\nonumber \\
 & = & 2\frac{\alpha}{D_{0}}\frac{z_{0}}{D_{0}}\frac{N}{j^{2}v_{2}^{2}}.\label{eq:delE_avg}
\end{eqnarray}

We now perform the integral over the impact parameter. The forgoing
equations can be extended to include all impact parameters by integrating
it with respect to the measure $D_{0}dD_{0}$. This corresponds to
taking into account all collisions where $m_{2}$ has $\left(x_{0},y_{0}\right)$
such that $D_{0}=\sqrt{x_{0}^{2}+y_{0}^{2}}$. The effect of rotations
in the $xy$ plane has been taken care of by the integral over $\theta$. 

\[
\langle\Delta E\rangle_{D_{0},\boldsymbol{v_{1}}}=\frac{2\alpha z_{0}N}{j^{2}v_{2}^{2}}\int\frac{dD_{0}}{D_{0}}=\frac{2\alpha z_{0}N}{j^{2}v_{2}^{2}}\log\left(\frac{D_{max}}{D_{min}}\right).
\]
Clearly the integral diverges for $D_{max}\rightarrow\infty$; this
is natural as the force is long range and by definition very distant
encounters need to be taken into account. It also diverges for $D_{min}\rightarrow0$,
which violates the distant encounter assumptions $\frac{\alpha}{\mu DV^{2}}\ll1$.
However, as it has been discussed in the context of astrophysics and
plasma physics (see Section \ref{sub:Effective-Collision-Times} and
\cite[chapter 2, pp. 55-57]{chandra4}) there is a natural $D_{min}$
that ensures small angle scattering and a $D_{max}$, depending on
the density of field particles, that appropriately characterizes the
maximum distant encounters. Further a factor of $2$ or $3$ error
in choosing $D_{max}$ does not affect the calculation of relaxation
times by much.

\begin{thebibliography}{10}
\bibitem{remark1} Interactions with $k$ smaller than the dimension
of space are considered long range.
\end{thebibliography}
\end{document}